\pdfoutput=1

\documentclass[fleqn,usenatbib]{mnras}
\usepackage{amsmath}
\usepackage{xspace}
\usepackage{mathptmx}
\usepackage{txfonts}
\usepackage{xcolor}

\usepackage[T1]{fontenc}
\usepackage{ae,aecompl}

\usepackage{graphicx}	
\usepackage{amssymb}	
\usepackage{booktabs}
\usepackage{wasysym}
\usepackage[normalem]{ulem}

\let\oldhref\href
\renewcommand{\href}[2]{\oldhref{#1}{\hbox{#2}}}

\definecolor{colorl1}{RGB}{0, 51, 153}
\definecolor{colorl2}{RGB}{153, 0, 0}
\definecolor{colorl3}{RGB}{179, 179, 0}
\definecolor{colorl4}{RGB}{51, 102, 0}

\definecolor{colorw1}{RGB}{51, 102, 255}
\definecolor{colorw2}{RGB}{255, 51, 0}
\definecolor{colorw3}{RGB}{255, 214, 51}
\definecolor{colorw4}{RGB}{51, 204, 51}

\newcommand{\hMpc}{{\ifmmode{h^{-1}{\rm Mpc}}\else{$h^{-1}$Mpc}\fi}}
\newcommand{\Mpc}{{\ifmmode{{\rm Mpc}}\else{Mpc}\fi}}
\newcommand{\hkpc}{{\ifmmode{h^{-1}{\rm kpc}}\else{$h^{-1}$kpc}\fi}}
\newcommand{\kpc}{{\ifmmode{ {\rm kpc} }\else{{\rm kpc}}\fi}}
\newcommand{\kms}{{\ifmmode{ {\rm km\,s^{-1}} }\else{ ${\rm km\,s^{-1}}$ }\fi}}
\newcommand{\hMsun}{{\ifmmode{h^{-1}{\rm {M_{\astrosun}}}}\else{$h^{-1}{\rm{M_{\astrosun}}}$}\fi}}
\newcommand{\Msun}{{\ifmmode{{\rm M}_{\astrosun}}\else{${\rm M}_{\astrosun}$}\fi}}
                        
\newcommand{\Mhalo}{{\ifmmode{M_{\rm halo}}\else{$M_{\rm halo}$}\fi}}
\newcommand{\Rvir}{{\ifmmode{R_{\rm vir}}\else{$R_{\rm vir}$}\fi}}
\newcommand{\Mvir}{{\ifmmode{M_{\rm vir}}\else{$M_{\rm vir}$}\fi}}
\newcommand{\Mstar}{{\ifmmode{M_{\rm star}}\else{$M_{\rm star}$}\fi}}
\newcommand{\Vrot}{{\ifmmode{V_{\rm rot}}\else{$V_{\rm rot}$}\fi}}
\newcommand{\ltsima}{$\; \buildrel < \over \sim \;$}
\newcommand{\gtsima}{$\; \buildrel > \over \sim \;$}
\newcommand{\lsim}{\lower.5ex\hbox{\ltsima}}
\newcommand{\gsim}{\lower.5ex\hbox{\gtsima}}

\def\lesssim{\mathrel{\hbox{\rlap{\hbox{\lower4pt\hbox{$\sim$}}}\hbox{$<$}}}}
\def\gtrsim{\mathrel{\hbox{\rlap{\hbox{\lower4pt\hbox{$\sim$}}}\hbox{$>$}}}}

\newcommand{\beq}{\begin{equation}}
\newcommand{\eeq}{\end{equation}}
\def\beqa{\begin{eqnarray}}
\def\eeqa{\end{eqnarray}}
\def\LCDM{\ensuremath{\Lambda}CDM}

\def\head{ \vbox to 0pt{\vss \hbox to 0pt{\hskip 440pt\rm
      LA-UR-10-07069\hss} \vskip 25pt}}

\def \kms {\ifmmode  \,\rm km\,s^{-1} \else $\,\rm km\,s^{-1}  $ \fi }
\def \kpc {\ifmmode  {\,\rm kpc}  \else ${\rm  kpc}$ \fi  }  
\def \hkpc {\ifmmode  {h^{-1}\rm kpc}  \else ${h^{-1}\rm kpc}$ \fi  }  
\def \hMpc {\ifmmode  {h^{-1}\rm Mpc}  \else ${h^{-1}\rm Mpc}$ \fi  }  
\def \Mpch {\ifmmode  {h^{-1}\rm Mpc}  \else ${h^{-1}\rm Mpc}$ \fi  }  
\def \Msun {\ifmmode {\rm M}_{\astrosun} \else ${\rm M}_{\astrosun}$ \fi} 
\def \hMsun {\ifmmode h^{-1}\,\rm M_{\astrosun} \else $h^{-1}\,\rm M_{\astrosun}$ \fi}
\def \Gyr {\ifmmode\, \rm Gyr \else $\,$Gyr \fi}

\def \LCDM {\ifmmode \Lambda{\rm CDM} \else $\Lambda{\rm CDM}$ \fi}
\def \sig8 {\ifmmode \sigma_8 \else $\sigma_8$ \fi} 
\def \OmegaM {\ifmmode \Omega_{\rm m} \else $\Omega_{\rm m}$ \fi} 
\def \Omegab {\ifmmode \Omega_{\rm b} \else $\Omega_{\rm b}$ \fi} 
\def \OmegaL {\ifmmode \Omega_{\rm \Lambda} \else $\Omega_{\rm \Lambda}$\fi} 
\def \Deltavir {\ifmmode \Delta_{\rm vir} \else $\Delta_{\rm vir}$ \fi}
\def \rhocrit {\ifmmode \rho_{\rm crit} \else $\rho_{\rm crit}$ \fi}
\def \rhou {\ifmmode \rho_{\rm u} \else $\rho_{\rm u}$ \fi}
\def \zc {\ifmmode z_{\rm c} \else $z_{\rm c}$ \fi}

\def\Mstar {\ensuremath {M_{*}(<r_{23.5})}~}
\def\r23_5 {\ensuremath {r_{23.5}}~}

\title[Protoplanet Collisions: Statistical Properties of Ejecta] {Protoplanet Collisions: Statistical Properties of Ejecta}

\author[Samuele Crespi et al.]{Samuele Crespi$^{1,2}$\thanks{E-mail: sc6459@nyu.edu},
Ian Dobbs-Dixon$^{1,2,3}$,
Nikolaos Georgakarakos$^{1,2}$,\newauthor
Nader Haghighipour$^{4,5}$, 
Thomas I. Maindl$^{6,7}$,
Christoph M. Schäfer$^{8}$, \newauthor
Philip Matthias Winter$^{9}$.
\\
\\
$^{1}$New York University Abu Dhabi, PO Box 129188, Abu Dhabi, United Arab Emirates \\
$^2$Center for Astro, Particle and Planetary Physics (CAP$^3$), New York University Abu Dhabi\\
$^3$Center for Space Science, New York University Abu Dhabi\\
$^4$Planetary Science Institute,
1700 East Fort Lowell, Tucson, AZ 85719, USA\\
$^5$Institute for Astronomy, University of Hawaii-Manoa, Honolulu, HI 96822, USA\\
$^6$Department of Astrophysics, University of Vienna, 1180 Vienna, Austria\\
$^7$SDB Science-driven Business Ltd, 6025 Larnaca, Cyprus\\
$^8$Institut für Astronomie und Astrophysik, Eberhard Karls Universität Tübingen, Auf der Morgenstelle 10, 72076 Tübingen, Germany\\
$^9$Institute for Machine Learning, Johannes Kepler University Linz, Altenberger Straße 69, 4040 Linz, Austria
}

\date{Accepted XXX. Received YYY; in original form ZZZ}

\pubyear{2021}

\begin{document}

\label{firstpage}
\pagerange{\pageref{firstpage}--\pageref{lastpage}}
\maketitle
\begin{abstract}

The last phase of the formation of rocky planets is dominated by collisions among Moon- to Mars-sized planetary embryos. Simulations of this phase need to handle the difficulty of including the post-impact material without saturating the numerical integrator. A common approach is to include the collision-generated material by clustering it into few bodies with the same mass and uniformly scattering them around the collision point. However, this approach oversimplifies the properties of the collision material by neglecting features that can play important roles in the final structure and composition of the system. In this study, we present a statistical analysis of the orbital architecture, mass, and size distributions of the material generated through embryo-embryo collisions and show how they can be used to develop a model that can be directly incorporated into the numerical integrations. For instance, results of our analysis indicate that the masses of the fragments follow an exponential distribution with an exponent of $-2.21\pm0.17$ over the range of $10^{-7}$ to $2\times 10^{-2}$ Earth-masses. The distribution of the post-impact velocities show that a large number of fragments are scattered toward the central star. The latter is a new finding that may be quite relevant to the delivery of material from the outer regions of the asteroid belt to the accretion zones of terrestrial planets.
Finally, we present an analytical model for the 2D distribution of fragments that can be directly incorporated into numerical integrations. 

\end{abstract}

\begin{keywords}
planets and satellites: formation -- planets and satellites: terrestrial planets
\end{keywords}

\section{Introduction}\label{sec:introduction}

Based on the standard model for planet formation, the last phase of rocky planet formation is dominated by collisions between Moon- to Mars-sized solid bodies, known as protoplanets or planetary embryos.
To a first approximation, this phase can be simulated as the evolution of a system of objects that interact exclusively via gravity (e.g., \citealt{Ida1993}, \citealt{Raymond2006}, \citealt{Barnes2009}, \citealt{Morishima2015}, \citealt{Clement2019}).
These simulations are generally carried out using N-body integrators with the largest constraint being the number of bodies, usually kept below a few hundred to avoid lengthy computations.

As the time of N-body integrations increases with the number of bodies, in order to ensure that formation simulations are carried out in a reasonable amount of time, collisions are often treated as perfectly inelastic. That is, the two colliding bodies merge completely and no fragments and debris are produced (e.g., \citealt{Raymond2004}, \citealt{OBrien2006} \citealt{Morishima2015}).
This approach has been proven to be useful in demonstrating the formation process, especially as a proof of concept. However, it tends to overestimate the collision efficiency, leading to shorter formation time and more massive planets. The latter causes the final planetary system to have smaller number of bodies with planets being farther away from the central star (\citealt{Doguaro2020}, \citealt{Burger2020}).

While simulating collisions has been an integral part of planetary science for decades (e.g., \citealt{1977Dormand}, \citealt{Benz1986}, \citealt{Wetherill1988}), inclusion of collisional fragments in formation simulations has not been straightforward. The reason lies in the fact that collisions produce million and millions of fragments, and including all these in an N-body integration is impractical. Many researchers have tried to circumvent this issue by including only a few of the impact-generated bodies (\citealt{Chambers2013}, \citealt{Clement2019}). Another approach is to use a collision catalogue to identify the post-impact bodies that have the larger contributions (i.e., those with larger mass) and ignore the rest (\citealt{Burger2020}). 

Although the above approach has advanced the simulations of terrestrial planet formation closer to realistic ones, it includes only a small portion of the impact-generated bodies. A  comprehensive formation model requires all this material to be included. However, as explained above, it is impractical to include all post-impact fragments as individual objects. Fortunately, the total mass of these bodies presents a promising path forward.

Because the effect of post-collision objects on the rest of the system is through their gravitational interactions and accretion, using the collective mass of the objects produced after a collision allows for both these factors to be taken into account. The only question will be, how to model this mass. \citet{Clement2019} suggested to distribute the total post-impact mass as a collection of lunar-mass bodies. \cite{Carter2015} included them as unresolved debris by uniformly distributing fragments into circular rings. While both these approaches have been promising, their underlying assumptions are too simplistic. In real collisions, fragments vary considerably in size and velocity, and their distribution is not uniform. A comprehensive model needs to take all this into account. In this paper, 
we investigate the statistical properties of the material ejected by collisions during the collision dominated phase of the rocky planet formation.
In particular, we will use a catalogue of SPH (Smoothed-Particles Hydrodynamics) simulations of collisions to determine the distributions of the mass, velocity, and orbits of the ejecta.

This paper is arranged as follows. We present in section \ref{sec:2} the two catalogues used in this study, namely the catalogue of N-body impacts and the catalogue of SPH impact simulations. In section \ref{sec:results}, we present and analyze the results of our simulations, and in section \ref{sec:conclusions}, we conclude this study by summarizing the results and dicsussing their implications.

\section{Collision Catalogues} \label{sec:2}

In this section, we present the two collision catalogues used in our study. The first catalogue consists of 1356 protoplanet collisions and is obtained by N-body simulations of the late stage of terrestrial planet formation in 10 different systems.  This catalogue, hereafter referred to as the PC (protoplanetary collision) catalogue, contains the physical properties of the colliding bodies prior to the time of each collision. The second catalogue contains the outcome of the SPH simulations of protoplanetary bodies for different initial conditions. We refer to this catalogue as CO (collision outcome) catalogue. In the following, we explain these two catalogues in more detail. 

\subsection{Protoplanets Collisions Catalogue}\label{PC}

To generate a collection of protoplanetary collisions, we simulated the late stage of terrestrial planet formation for 10 different distributions of planetary embryos. We considered a system consisting of the Sun, Jupiter in a circular orbit at 5.2 AU, and a disk of protoplanetary bodies extending from 0.5 AU to 4.7 AU. The protoplanetary disk contained around 200 planetary embryos with masses ranging from 0.02 to 0.1 Earth-masses. Following \citep{Kokubo2000}, planetary embryos were placed at distances of 5-10 mutual Hill's radii, and the surface density of the disk was set to followed a $\Sigma=\Sigma_1 r^{-3/2}$ profile with $\Sigma_1=10$ g/cm$^3$. The 10 different simulations differed only in the initial mass and orbits of protoplanets. We integrated each system for 1 Myr using the hybrid routine in the N-body integration package {\it Mercury} \citep{Chambers1999}. The integration time step was set to 6 days.

The collisions in the PC catalogue are uniquely identified by three parameters: the total mass of the colliding bodies (${m_1} + {m_2}=m_\mathrm{tot}$), their mass-ratio $(\gamma=m_1/m_2)$, and their impact velocity ($v_0$) normalized to the mutual escape velocity (hereafter, normalised impact velocity)\footnote{The mutual escape velocity is given by  \mbox{$v_{\mathrm{esc}}=[{2G(m_1+m_2)/(r_1+r_2)}]^{1/2}$}, where $r_1$ and $r_2$ are the radii of the two colliding bodies and $G$ is the gravitational constant.}.
Here, $m_1$ and $m_2$ are the masses of the two colliding bodies at the moment of impact.
Other parameters, such as the collision angle, can also play a role in defining the collisional outcome.
However, to avoid computational complexities, they are not investigated in this study.
Instead, we consider the averaged value of these parameters, as described in Section \ref{Interp_explanation}.

\begin{figure}
  \hspace*{-5mm}
    \includegraphics[width=0.54\textwidth]{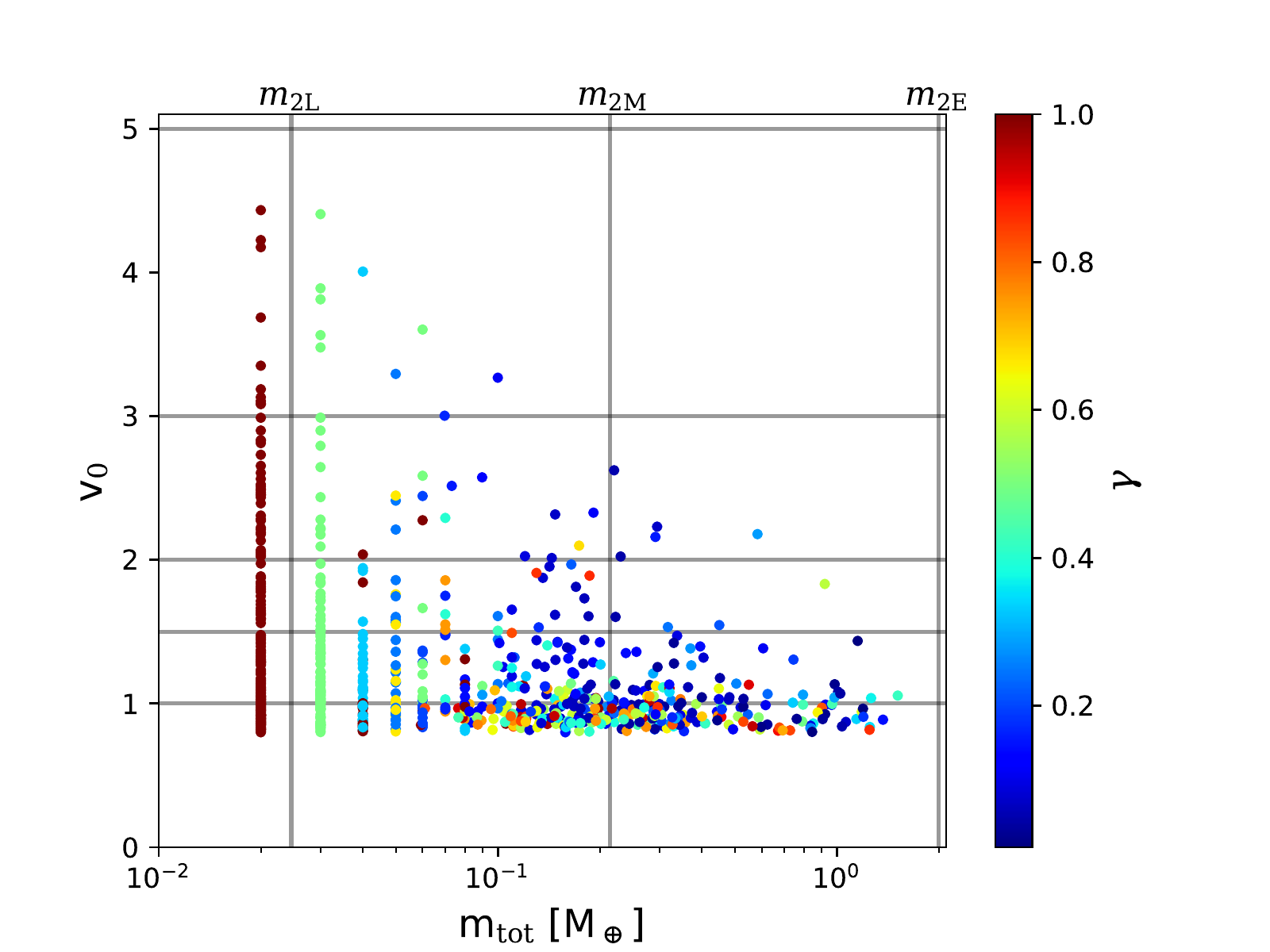}
    \caption{Distribution of all 1356 collisions in the PC catalogue. The $x$ and $y$ axes show the total mass and the normalised impact velocity, respectively. The color-coding indicates the mass-ratio of the colliding bodies. The horizontal and vertical lines correspond to the specific set of values sampled in the CO catalogue. }
    \label{m_gamma_v0}
\end{figure}

\begin{figure}
  \vspace*{5mm}
  \hspace*{0mm}
    \includegraphics[width=0.45\textwidth]{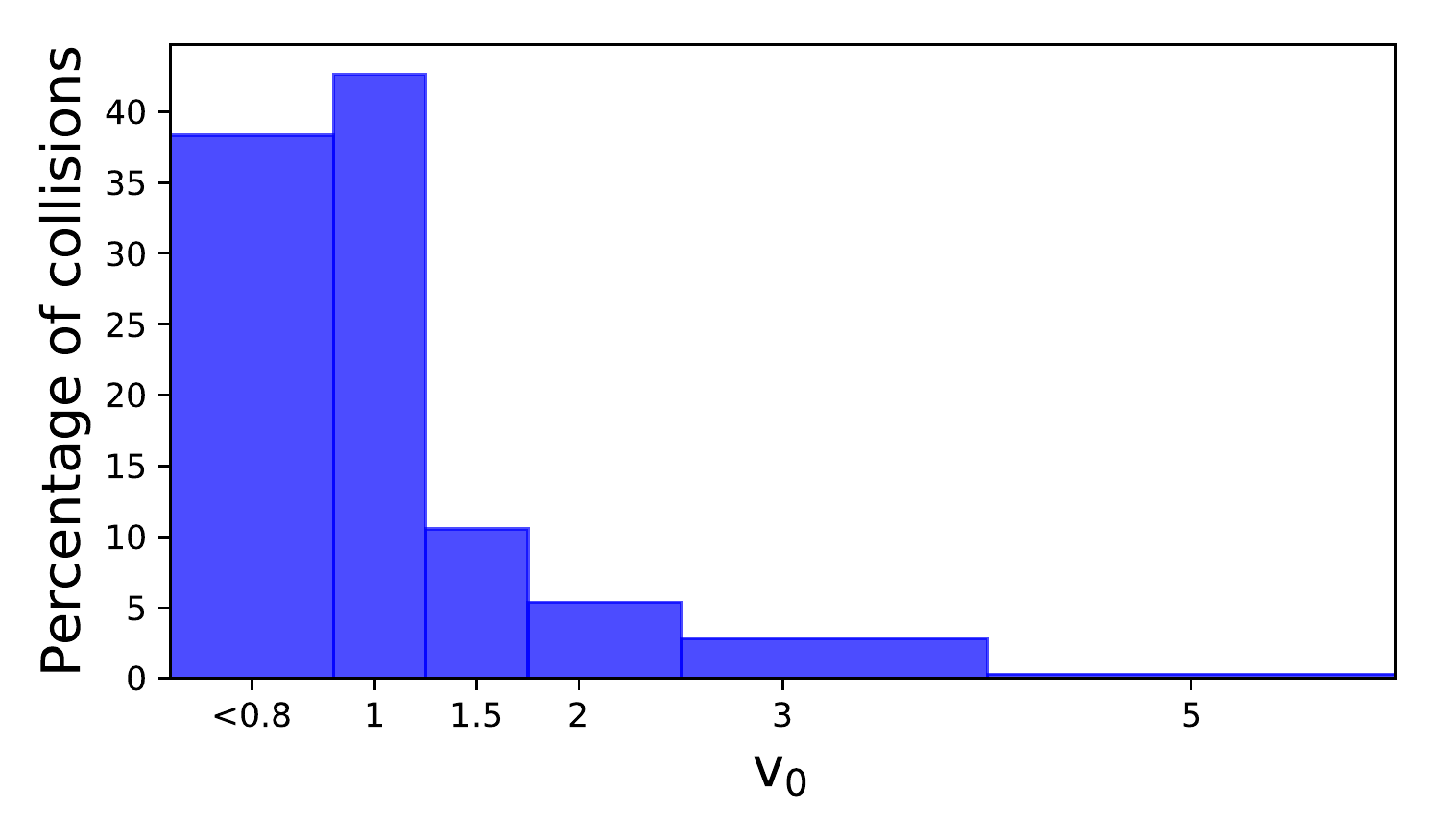}
    \caption{Histogram of the normalised impact velocity $(v_0)$ for the collisions of the PC catalogue. 
    }
    \label{v0_gamma_histos}
\end{figure}

Figure \ref{m_gamma_v0} shows the distribution of all 1356 collisions in term of $m_\mathrm{tot}$, $\gamma$ and $v_0$.
As shown here, high velocity collisions ($v_0>3$) are extremely rare (less than 2\% of all collisions) and usually occur among smaller bodies ($m_{\rm tot}<7\times 10^{-2}$ M$_\oplus$). The dominance of low velocity impacts is more evident in Figure \ref{v0_gamma_histos} where we show a histogram of collision velocities. As many as 893 collisions (approximately 2/3 of all collisions) have impact velocities smaller than the mutual escape velocities of the colliding bodies ($v_0<1$).

\subsection{Collision Output Catalogue}

Our catalogue of collision outcomes consists of 880 low-resolution SPH simulations (about 20000 particles\footnote{\cite{Burger2018} analysed collision outcomes for resolutions between a few 10k and 2.25M SPH particles. They found that even for highly energetic collisions of Earth- and Mars-mass bodies, the global outcomes (kinematics and masses of fragments) show only minor deviations below 10\% with varying resolutions, hence only marginally impacting the global outcome classification. The water mass fraction was found to be accurate within only a few percent for the largest surviving body and within about 25\% for the second largest fragment. Being mainly interested in the overall surviving fragment masses and kinematics, we settle for a resolution of 20k SPH particles to balance performance and accuracy.} per scenario) of two colliding protoplanets.
The collision simulations were performed using the GPU SPH code \texttt{miluphcuda} 
(\citealt{Schafer2016}, \citealt{Schafer2020}). 

The initial setup of projectile and target includes SPH particles on hexagonal close-packed grids in hydrostatic equilibrium. Target and projectile are initially set as not rotating and placed at a mutual distance of $r_\mathrm{ini} = 5 (r_1 + r_2)$, where $r_1$ and $r_2$ denote the radius of the target and projectile, respectively.

Both the projectile and target are modelled as silicates surrounded by ice shells to account for their water-mass fraction. We use the Tillotson equation of state with input parameters of basalt and ice \citep{Schafer2016} to determine the pressure. The material strength is also included where the shear strength is limited by the von Mises yield criterion. To model tensile fractures, we follow \cite{Benz1994} and use the Grady-Kipp fragmentation model and the usual standard artificial viscosity by \cite{Monaghan1985}. The gravitational forces between the particles are calculated using a Barnes-Hut tree.

Table \ref{SPH_params} shows the range of initial conditions for the collision simulations.
The values of the mass correspond to the total colliding mass of two Ceres-sized bodies (hereafter labelled $m_\mathrm{2C}$), two Moon-sized bodies (labelled $m_\mathrm{2L}$), two Mars-sized bodies (labelled $m_\mathrm{2M}$) and two Earth-sized bodies (labelled $m_\mathrm{2E}$).
To maintain generality, two different values of water-mass fraction are considered for each body. However, the composition of the ejected material is not investigated.

\subsubsection{Collision-Type Criteria}\label{Coll_type_crit}

Each SPH simulation produces the positions, velocities and masses of all bodies formed $\sim$9 to 12 hours after the impact. To determine the type of the collision, the post-impact bodies must be examined
for possible gravitationally bound objects and also to separate the main bodies from fragments.

To determine if two bodies are gravitationally bound, we compare their relative velocity ($v_\mathrm{rel}$) with their mutual escape velocity. 
If $v_\mathrm{rel}$ is smaller than the mutual escape velocity, we consider the two bodies gravitationally bound and combine them into a single object at their center of mass. We continue this procedure until no pairs of gravitationally bound ejecta are left. Results indicated that 15\%-30\% of the post-impact material are gravitationally bound to their main bodies and is accreted back onto those objects.

To separate the main bodies (meaning the remnants of the colliding bodies if not destroyed) from
fragments, we use the following criterion: if the impacting body loses more than 90\% of its original mass, it is considered to have been destroyed and is no longer a main body \citep{Leinhardt2012}. 

Through these two processes, we identify 5 possible types of collisions:
\begin{itemize}
    \item \textit{Perfect Merging} (PM): the two colliding bodies merge and produce one main body with no fragments;
    \item \textit{quasi-Perfect Merging} (qPM): more than 90\% of one of the two colliding bodies is accreted by the other producing one main body and a small number of fragments;
    \item \textit{Partial Accretion} (PA): less than 90\% of one of the two colliding bodies is accreted by the other resulting in one or two main bodies and fragments;
    \item \textit{Erosive Collision} (EC): the biggest body loses less than 90\% of its mass during the collision resulting in one or two main bodies and fragments;
    \item \textit{Catastrophic Collision} (CC): both colliding bodies are destroyed and all mass is converted into fragments.
\end{itemize}{}
We would like to note that the above five collision outcomes are rather different from those described by \cite{Leinhardt2012}. In particular, these authors divide PA and EC collisions into smaller categories including pure hit-and-run, erosive hit-and-run, cratering and grazing collisions. We decided not to follow this categorization as, in our dataset, it is impossible to track which bodies are the two colliding bodies at the end of a simulation.

\begin{table}
\centering
\begin{tabular}{l|l}
\hline
Parameter     & value \\
\hline
Total mass [M$_\mathrm{\oplus}$] & $3.14\times 10^{-4}$, $2.46\times 10^{-2}$,\\
 &$0.215$, 2\\
Mass-ratio  & 0.1, 0.5, 1.0 \\
Impact velocity [$v_{\mathrm{esc}}$] & 1, 1.5, 2, 3, 5 \\
Impact angle [deg] & 0, 20, 40, 60 \\
Body 1 water mass fraction $w_1$ & 0.1, 0.2 \\
Body 2 water mass fraction $w_2$ & 0.1, 0.2 \\
\hline
\end{tabular}
\caption{Parameter values sampled in the 880 simulations of the CO catalogue.}
\label{SPH_params}
\end{table}

\begin{figure}
\vspace{-0.6cm}
\hspace*{-0.3cm}
\includegraphics[width=0.53\textwidth]{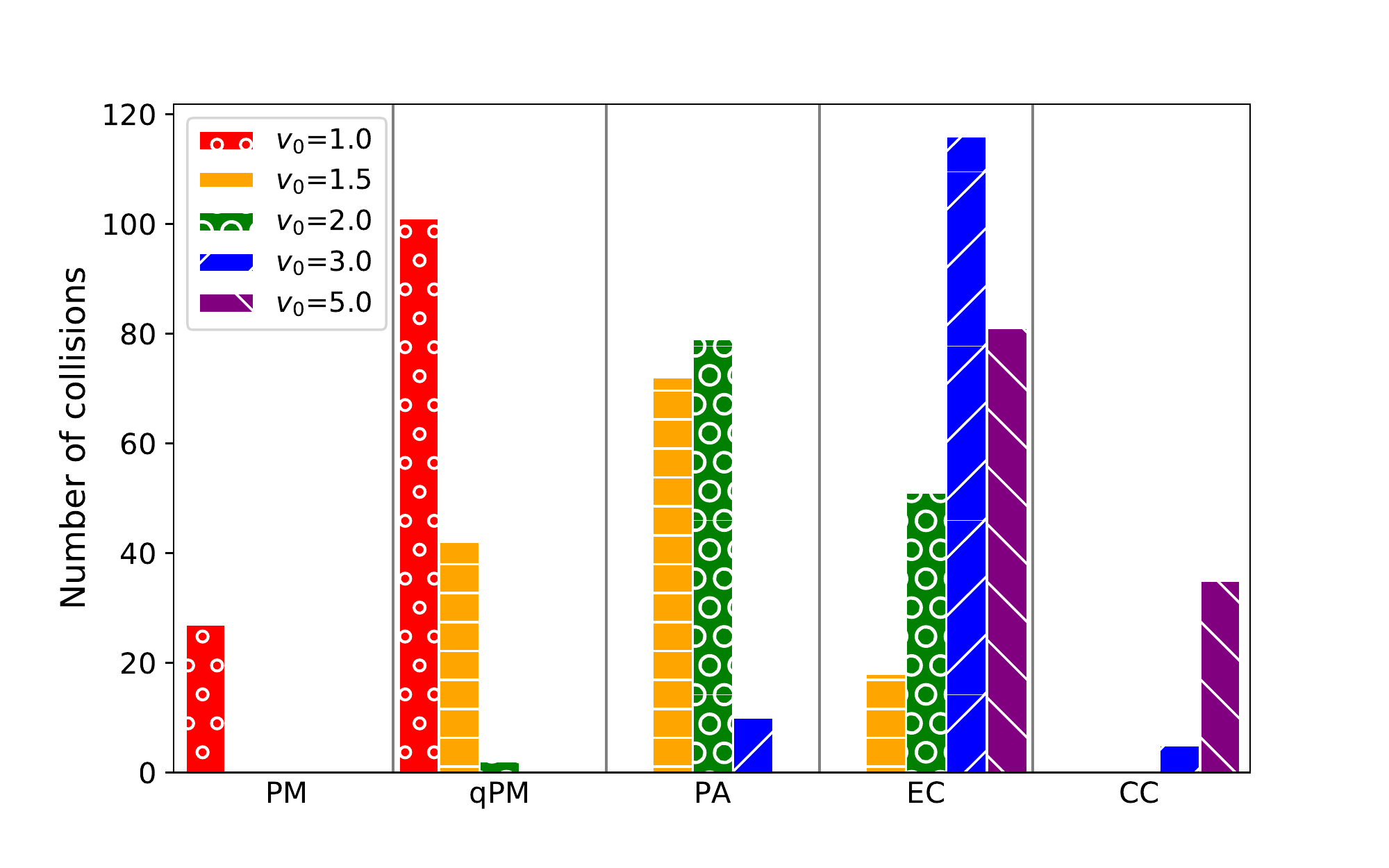}
\vspace{-0.6cm}
\caption{Occurrence of different collision outcomes in all the SPH simulations in the CO catalogue. The color coding shows the impact velocity in units of mutual escape velocity.}
\label{fig:coll_type}
\end{figure}

Results of our SPH simulations indicated that the collision type strongly depends on the impact velocity and the impact angle while the effect of the other parameters such as porosity and material composition is almost negligible.
In particular:

\begin{itemize}
    \item for collisions with impact velocities comparable with the escape velocity ($v_\mathrm{0}= 1$) the result is always a PM or qPM,
    \item collisions with impact velocities at least three times larger than the escape velocity ($v_\mathrm{0}> 3$) never result in PM,
    \item collisions with impact velocities five times larger than the escape velocity ($v_\mathrm{0}= 5$) either result in EC or CC.
\end{itemize}  
These findings are consistent with the predicted collision outcomes in \cite{Leinhardt2012}.
Figure \ref{fig:coll_type} shows the occurrence of these collisions as a function of the impact velocity $v_\mathrm{0}$ for all our SPH simulations.

\section{Analysis and Discussion}\label{sec:results}

In this section, we use the SPH results from the CO catalogue to present an analysis of the possible outcomes of the collisions in the PC catalogue. To begin with, we use the collision classification introduced in the last section to identify the type of collisions that can emerge from the PC catalogue. Figure 4 shows the results. As shown here, a large majority of the collisions (84.9\%) result in perfect or quasi-perfect merging. Other types of collisions also contribute, but at much smaller extent; partial accretion appears at 7.7\%, erosion occurs at 7.2\%, and complete shattering of the colliding bodies (catastrophic collision) happens only 0.18\% of the time.     

\subsection{Total Mass of Fragments}

Using the results of the SPH simulations, we determined the total mass of the fragments that could be produced in each PC collision. 
We found that impacts with partial accretion would have the biggest contribution, producing most of the fragments' mass. The largest post-impact total mass produced in all collisions in the PC catalogue is that of a PA collision between two embryos with masses of $0.29 M_\oplus$ and $0.26 M_\oplus$. Despite their relatively low impact velocity ($v_\mathrm{0}=1.8$), these bodies can produce approximately $0.1 M_\oplus$ of fragments, equivalent to 17\% of the initial mass of their corresponding system. Results indicate that on average, three such collisions appear in each formation simulation where the collective contribution of these collisions to the total mass of produced fragments could be as large as $\sim 0.067 {M_\oplus}$. 

Our analysis also indicates that although catastrophic collisions convert all colliding masses into debris, their contributions to the total mass of the fragments are small. Because these collisions happen only for high impact velocities ($v_\mathrm{0}>4$) and relatively small total colliding mass ($m_\mathrm{tot}<0.07$ M$_\oplus$), they are rare. Only 0.14\% of all collisions in the PC catalogue are of CC type. For that reason, these collisions do not have significant contributions to the total mass of fragments. 

Results showed that on average, in each simulation, \mbox{0.192 M$_\oplus$}, corresponding to 4.2\% of the total initial mass of a system is converted into debris. This value is smaller than those reported in the study by \citet{Burger2020} where 18\% -- 24\% of the initial mass is converted into fragments. We believe this discrepancy has roots in the differences between the initial setups in our simulations and those by these authors. For instance,
these authors included Saturn which can further perturb the orbits of planetary embryos and increase their impact velocities. They also included more bodies which increases the mutual interactions among the embryos, subsequently increasing their impact velocities. The 4.2\% total fragments mass in our simulations should, therefore, be taken as a conservative lower limit which is still significant enough to emphasize the importance of including fragments in formation simulations. The latter is especially important when studying the composition of the final bodies.

\begin{figure}
\vspace{-0.5cm}
\hspace*{-0.5cm}
\includegraphics[width=0.53\textwidth]{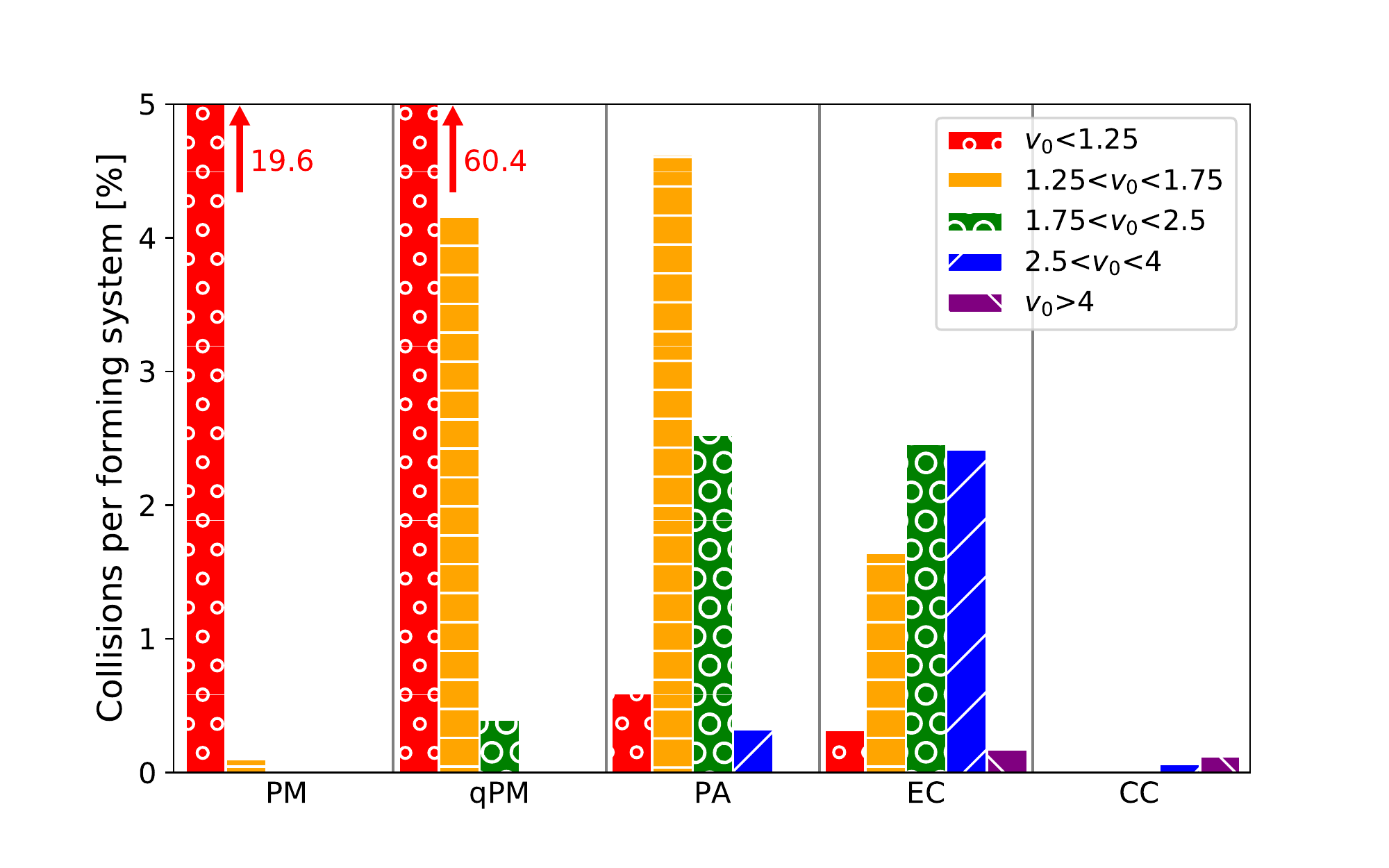}
\caption{Occurrence of the 5 different collision types per forming system (PC catalogue). The color and texture indicates the impact velocity interval in units of the mutual escape velocity between the two colliding bodies.}
\label{fig:PCcoll_type}
\vspace{-0.cm}
\end{figure}

We would like to note that for the sake of completeness and generality, we also carried out 10 more PC-type simulations where we included Saturn as well. We will analyze these collisions and the degree to which Saturn contributed to the collision outcomes in section \ref{sub_sec_a-e_distr}.

\subsection{Fragment Mass Distribution}

The mass distribution of the post-impact material is often neglected in simulations that include fragmentation. It has been customary to include all post-impact material into a few bodies, in some instances, even with equal masses, (e.g., \citealt{Clement2019,Poon2020,Scora2020}).
However, as pointed out by some authors (\citealt{Mustill2018,Doguaro2020}), fragments of different masses can affect the evolution of the system differently, both from a physical and a chemical perspective. In this section, we present an analysis of the mass distribution of post-impact bodies and we derive an equation that can be incorporated into formation simulations.

\begin{figure}
  \centering
    \includegraphics[width=0.47\textwidth]{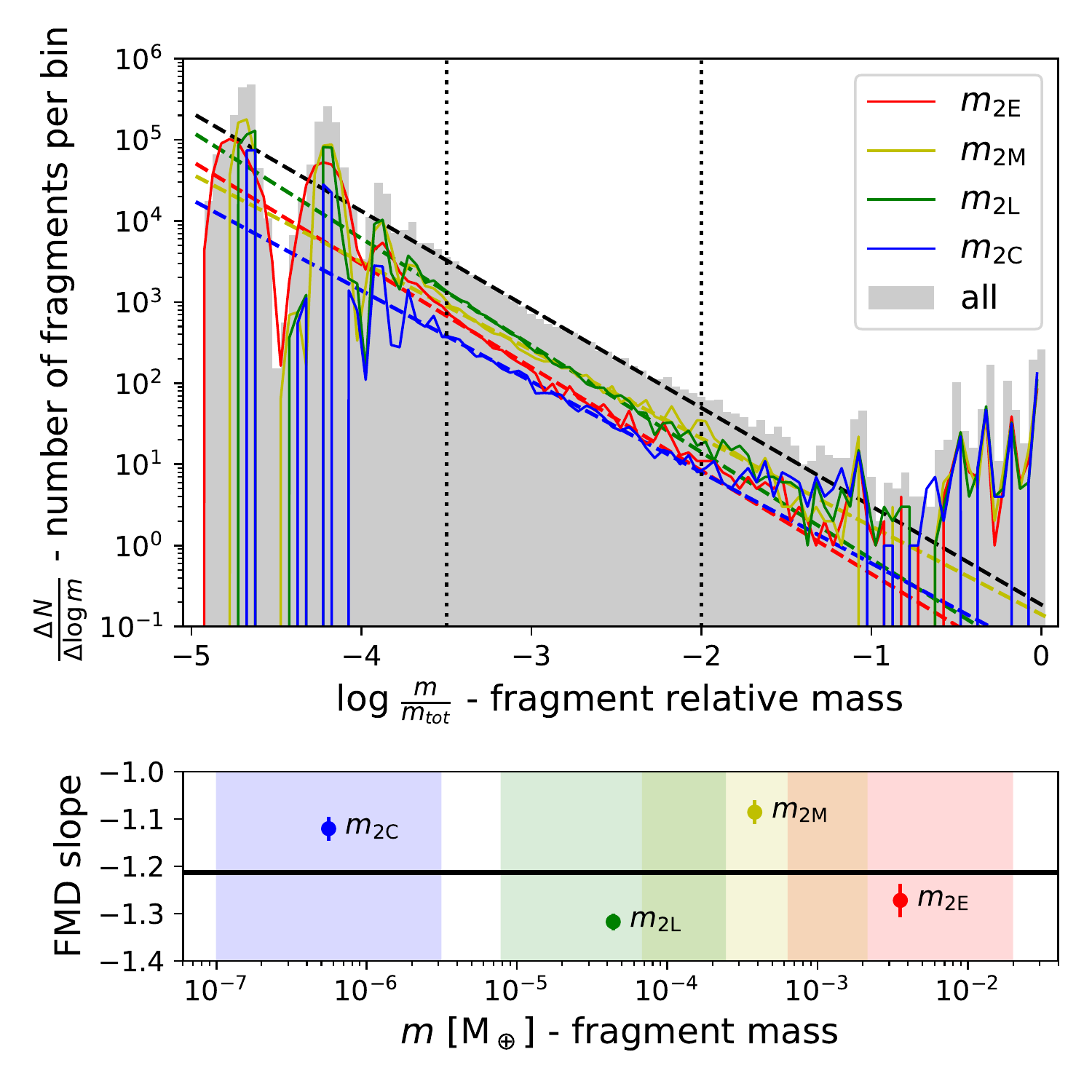}
  \caption{\textit{Top}: Normalised fragments mass distribution. The gray area corresponds to all the simulations of the CO catalogue. The four colored lines correspond to simulations with different total masses: red for $m_\mathrm{2E}$, yellow for $m_\mathrm{2M}$, green for $m_\mathrm{2L}$ and blue for $m_\mathrm{2C}$. The dashed lines indicate the best linear fit for each distribution in the range $10^{-3.5}$-$10^{-2}$ (i.e. within the two vertical dotted lines).
  \textit{Bottom}: Distribution of the slopes of fits in the top panel. The coloured areas correspond to the mass range used in each fit and their corresponding slope are shown at the center of each range. The black line indicates the slope of the overall distribution. }\label{FMD_h}
\end{figure}

Figure \ref{FMD_h} shows a histogram of the masses of the fragments in all simulations of the CO catalogue. The colored graphs correspond to the mass distribution of fragments produced in the four subsets of collisions with the total colliding masses of $m_\mathrm{2C}$, $m_\mathrm{2L}$, $m_\mathrm{2M}$, and $m_\mathrm{2E}$  as given in Table 1. As shown here, no significant differences exist in the mass distributions of these impacts. In all four cases, the region corresponding to $({10^{-3.5}} - {10^{-2}})\,{m_{\rm tot}}$ shows a linear trend in log-log space with a slope varying from -1.09 for the $m_\mathrm{2M}$ collision to -1.32 for the system of $m_\mathrm{2L}$. Figure 5 also shows that except for the small interval of ${(3.1-7.8) 10^{-6}} M_\oplus$ (which is due to the fact that $m_\mathrm{2C}$ masses are two orders of magnitude smaller than those of $m_\mathrm{2L}$), the distribution of final fragments masses show an overlap (see the bottom panel). The latter suggests that, in the range of $({10^{-3.5}} - {10^{-2}})\,{m_{\rm tot}}$, the post-impact distribution of mass in all these four collisions can be studied collectively. We found that, together, these fragments present an exponential distribution given by

\begin{equation}\label{FMD}
n(m)\,\mathrm{d}m \propto m^{-2.21\pm0.17}\,\mathrm{d}m\,.
\end{equation}
Here, $n(m)$ is the number of fragments with mass between $m$ and $m+\mathrm{d}m$. It is important to mention that the choice of the mass-range of $({10^{-3.5}} - {10^{-2}})\,{m_{\rm tot}}$ was to prevent the effects of computational uncertainties. For instance, for smaller values of fragments' masses, the distribution strongly depends on the resolution of the SPH simulations, and for larger values, the distribution is dominated by the largest survivors.
In particular, the 5 peaks between -1.04 and 0 correspond to hit-and-run scenarios for which the mass of two colliding bodies remain nearly unchanged after the collision.

In order to obtain the error on the slope value, we performed a similar analysis for sets of collisions grouped according to the mass ratio $\gamma$ and the normalised impact velocity $v_\mathrm{0}$.
We found that the slope of the mass distribution varies between -2.58 and -1.96 with a standard deviation of 0.17 for collisions with $v_\mathrm{0} = 5$  and 3, respectively.
We also noticed that as the mass ratio $\gamma$ increases from 0.1 to 0.5 and 1, the absolute value of the slope of the mass distribution decreases from -2.51 to -2.22 and -2.12.

Assuming constant bulk density and spherical shape for all fragments, equation \ref{FMD} can be written as
\begin{equation}\label{FMD_D}
    N(D)\,\mathrm{d}D \propto D^{-4.63\pm 0.52}\,\mathrm{d}D\,,
\end{equation} 
where $D$ represents a fragment's diameter and $n(D)$ is the number of fragments with size between $D$ and $D+dD$. The distribution slope, that in the log-log space becomes $-3.63\pm 0.52$, is in full agreement with the median slope of -3.8 reported by \cite{Leinhardt2012}.
Equations \ref{FMD} and \ref{FMD_D} can be used to model fragmentation more realistically and to improve the mechanics of the inclusion of fragments in formation simulations.

\subsection{Distribution of Fragments}\label{Interp_explanation}

In this section, we determine the distribution of the fragments that can be produced in the collisions among planetary embryos. Because N-body integrations do not resolve embryo-embryo collisions, we use the orbital parameters of colliding bodies at the moment of their impact (i.e., their position in the parameter-space of the PC catalogue) to identify their possible outcome using the results of the SPH simulations (i.e., their location in the parameter-space of the CO catalogue).

In general, for each N-body collision, the knowledge of the total colliding mass ($m_\mathrm{tot}$), mass-ratio ($\gamma$), and normalised impact velocity ($v_\mathrm{0}$) would be sufficient to uniquely identify that collision in the PC catalogue. However, because the two catalogues are independent from each other, each parameter set $({m_\mathrm{tot}}, \gamma, {v_\mathrm{0}})$ from the PC catalogue may not necessarily correspond to an exact initial condition for an SPH simulation in the CO catalogue. It is, therefore, necessary to develop a methodology to determine the degree to which the results of an SPH simulation can be used for a given set of embryo-embryo collision. This can be done by weighting the difference between the initial condition of an embryo-embryo collision in the PC catalogue and its closest initial condition in the parameter-space of the CO catalogue. In the following, we explain our approach in developing such a weighting measure.

Recall that the initial conditions of the SPH simulations were divided into grids along all parameters. Let's assume that a parameter $x$ from a set of initial conditions of a collision in the PC catalogue falls in the $i$-th grid in the parameter-space of the CO catalogue where the value of the same quantity is given by $x_i$. We introduce the function $K$, hereafter referred to as the weight function, as a measure of the difference between the values $x$ and $x_i$,

\begin{equation}\label{kernel}
    K(d) = \begin{cases}
    1 & \mathrm{if}\,\,\,d_i<\delta\\
    2-(d/{\delta}) & \mathrm{if}\,\,\,\delta<d_i<2\delta \,.\\
    0 & \mathrm{if}\,\,\,d_i>2\delta
    \end{cases}
\end{equation}
Here, $d_i=\mid x - x_i\mid$. We define the quantity \mbox{$\delta=(x_i-x_{i+1})/3$} as the scaling length of the weight function corresponding to one-third of the difference between two consecutive intervals of the quantity $x$ in the parameter-space of the CO catalogue. We choose this finite-length because, as demonstrated by the results of the SPH simulations, the outcome of a collision can vary considerably from one grid point to the next. From equation \ref{kernel}, if for a parameter $x$ in the PC catalogue, $d_i>2\delta$, we consider that parameter to be too far from $x_i$ and, therefore, the results of the SPH simulations corresponding to the initial condition $x_i$ cannot be used for that embryo-embryo collision in the PC catalogue.

\begin{figure}
  \centering
  \hspace{-0.3cm} 
    \includegraphics[width=0.49\textwidth]{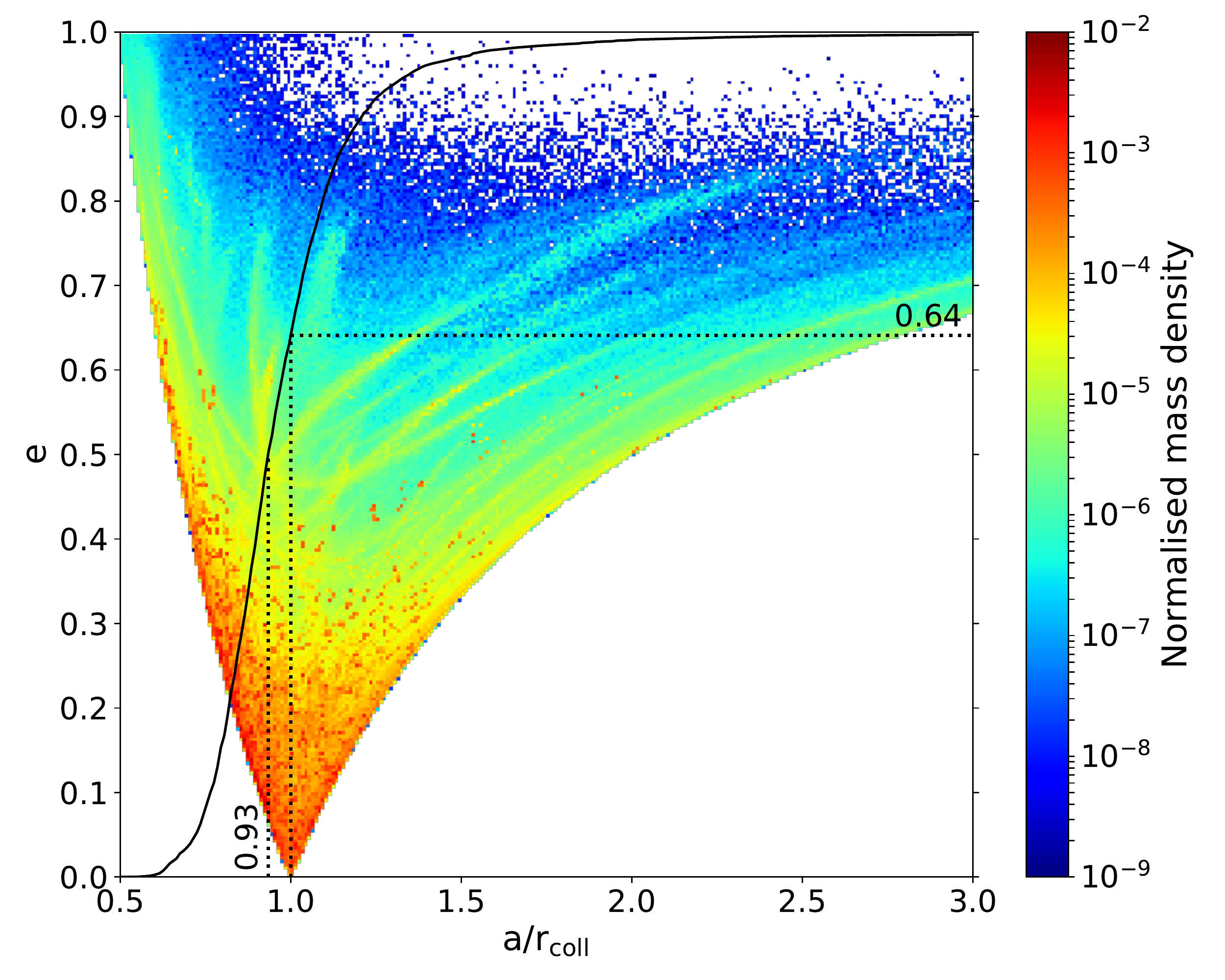}
  \caption{Fragments mass distribution in the $a-e$ plane. The semimajor axis is normalised by the radial distance between the star and the collision point $r_\mathrm{coll}$. The over-plotted black line corresponds to the normalised cumulative mass of the fragments and the dotted lines highlight the points at $a/r_\mathrm{coll}=1$ and \mbox{$m_\mathrm{cumulative}=0.5$}. The streaks represent fragments produced during high-velocity impacts ($v_\mathrm{0}>4$).}\label{dm_dade}
\end{figure}

\begin{figure}
  \centering
  \hspace*{-0.3cm}
    \includegraphics[width=0.49\textwidth]{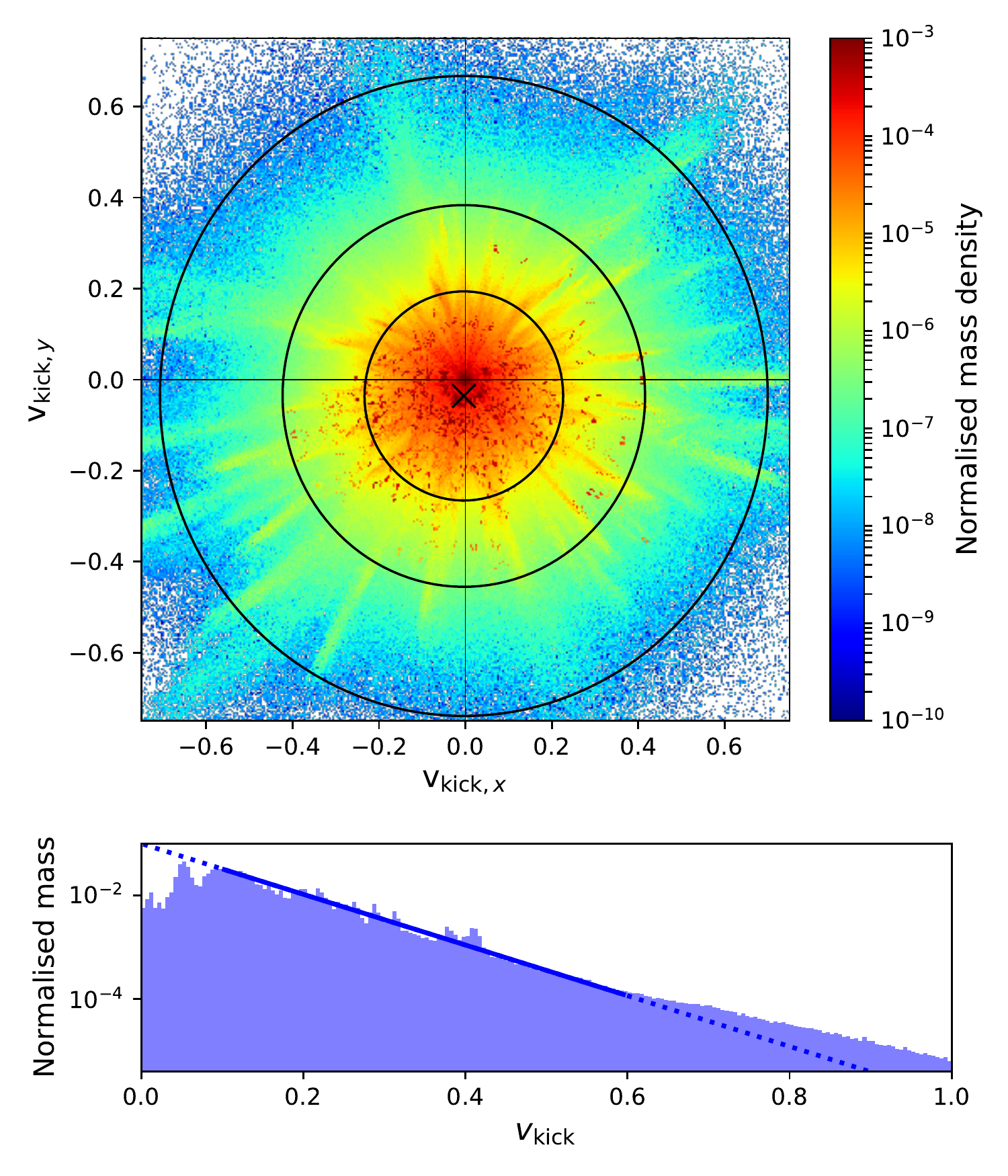}
  \vspace*{-5mm}
  \caption{\textit{Top}: Distribution of the relative velocity $v_\mathrm{rel}$ of fragments with respect to the center of mass of the system. The $x$ and $y$ axes are oriented such that for each collision, the velocity of the center of mass (shown by X) has components of $\mathbf{\hat{v}}_\mathrm{CoM}=(0,1)$ and the total orbital angular momentum of the system points towards the reader. The three concentric circles correspond to the regions containing 90\%, 99\% and 99.9\% of the total fragmented mass. \textit{Bottom}: Histogram of the distribution of fragments velocities. 
  The blue line corresponds to the exponential fit in equation \ref{kics_fit}. The dotted part of the line has been excluded from the fitting process.}\label{kicks}
\end{figure}

In applying equation \ref{kernel} to the results of the PC catalogue, it is important to note that for each set of $({m_\mathrm{tot}}, \gamma, {v_\mathrm{0}})$ parameters, SPH simulations were carried out for four different values of the impact angle and four different combinations of water-mass fraction. That means, for each set of the initial conditions in the PC catalogue, there are 16 SPH simulations with different impact angles and water content. Because the collisions in the PC catalogue do not include these parameters, we use the following averaging procedure to incorporate their total effects into account.

To remove the effects of water-mass fraction and impact angle for a given embryo-embryo collision, we calculate the total fragment mass $({m_{\rm fr}})$ of that collision by averaging over all its 16 values and determine its corresponding location in the CO parameter-space using equation \ref{kernel};

\begin{equation}\label{mfr_av}
{m_{\rm fr}}({m_{\rm tot}}, \gamma, {v_0}) = 
 \frac{1}{16\,{m_{{\rm tot}}}}\,{\sum_{i,j,k}} {K_i} {K_j} {K_k} {\sum_{l,m,n}} {m_{\rm fr}}
 ({m_{\rm tot}}, \gamma, {v_0}, {w_{1i}}, w_{2j}, {\alpha_k})
\end{equation}

 While this procedure removes information about the mass of individual fragments, it gives the probability for having a fragment with a specific mass and initial conditions $({m_\mathrm{tot}}, \gamma, {v_\mathrm{0}})$.

We apply equations \ref{kernel} and \ref{mfr_av} to all 1356 collisions in the PC catalogue. To ensure that for each embryo-embryo collision, the quantities representing lengths are compatible in both catalogues, we scale semi-major axis $(a)$ and radial distance $(r)$ in the PC catalogue by the distance of the location of the collision from the star $(r_\mathrm{coll})$. The final result consists of 9,373,158 fragments with minimum mass of $10^{-6}$M$_\oplus$ for which we know the normalised semi-major axis $(a/r_\mathrm{coll})$, eccentricity $(e)$, and argument of the periastron $(\omega)$. In the following, we use this approach to determine the distribution of post-impact material in the $a-e$ parameter-space, and in the plane of the system.

\begin{figure}
  \centering
  \hspace{-0.3cm} 
    \includegraphics[width=0.49\textwidth]{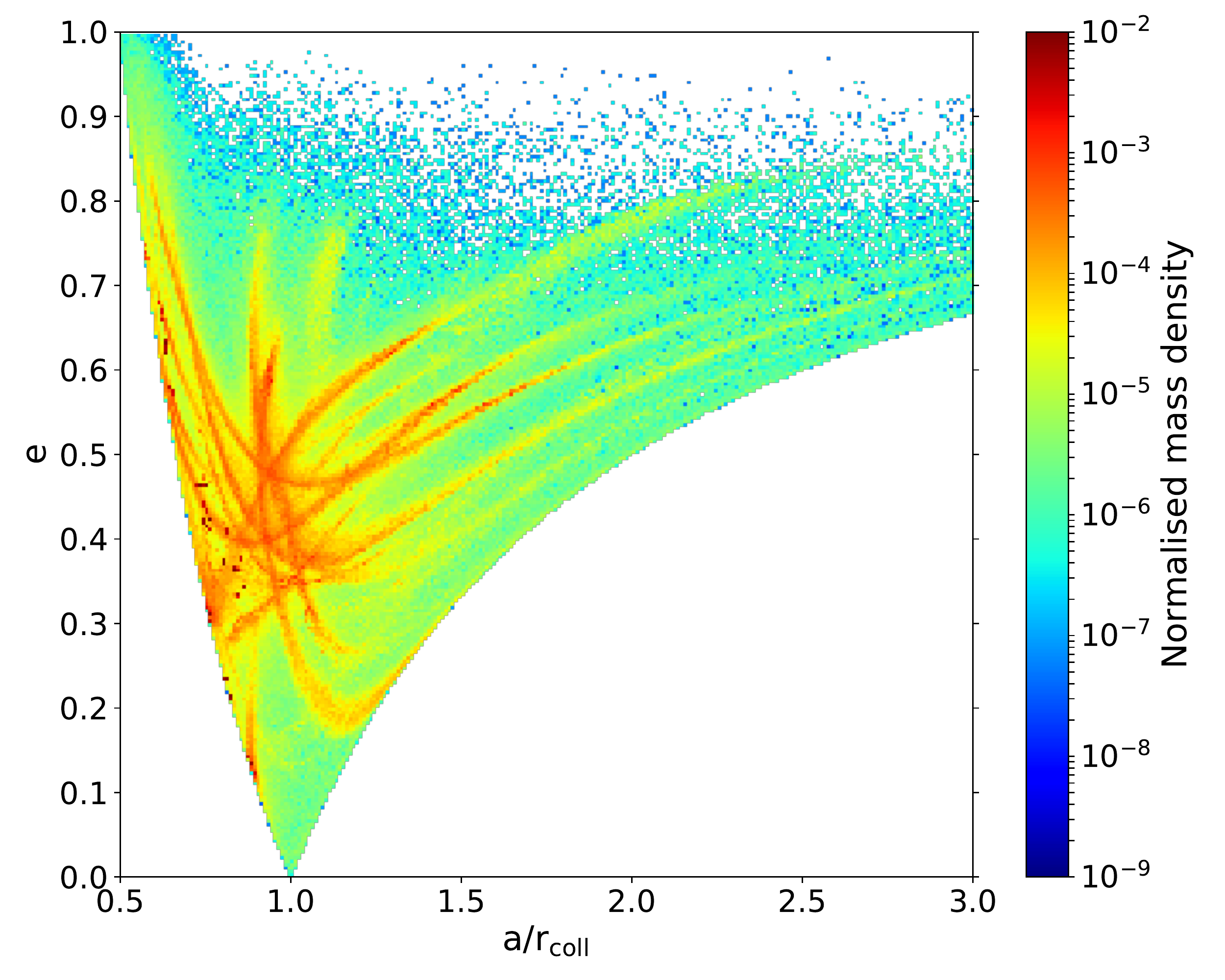}
  \caption{Similar to figure \ref{dm_dade} but including only the collisions with normalised impact velocity $v_\mathrm{0}>2$ and eccentricity $e_\mathrm{CoM}>0.33$.}\label{dm_dade_filtered}
\end{figure}

\begin{figure}
  \centering
  \hspace*{-4.7mm} 
    \includegraphics[width=0.5\textwidth]{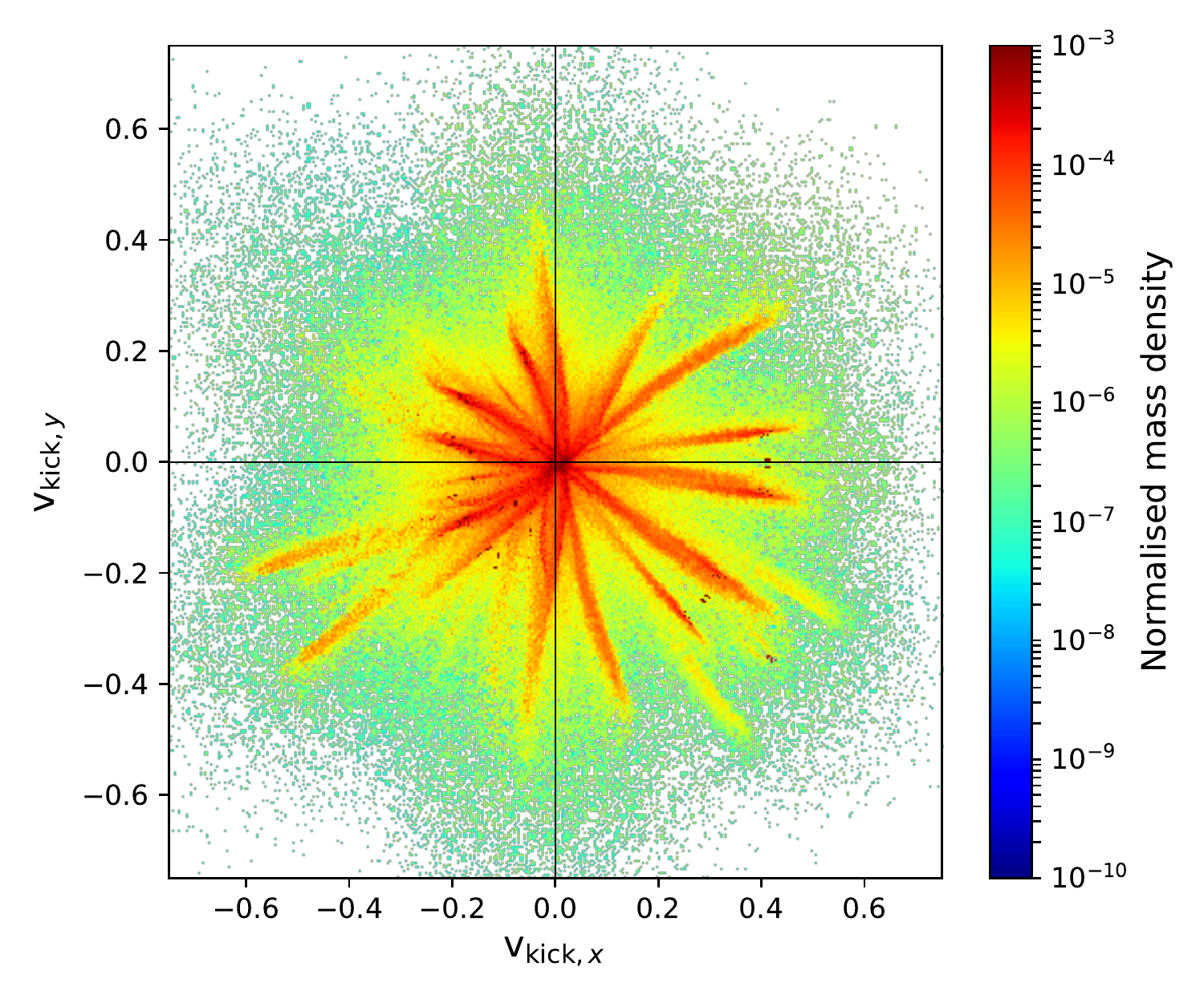}
  \vspace*{-5mm}
  \caption{Similar to figure \ref{kicks} but including only the collisions with normalised impact velocity $v_\mathrm{0}>2$ and eccentricity $e_\mathrm{CoM}>0.33$.}\label{kicks_filtered}
\end{figure}

\subsubsection{Fragments distribution on the $a-e$ plane}\label{sub_sec_a-e_distr}

Figures \ref{dm_dade} shows the distribution of fragments on the $a -e$ plane. The color-coding represents the amount of the mass confined in a grid-cell, per unit area of the cell. The solid black curve shows the cumulative normalized value of fragments mass. As expected, fragments are distributed between two branches \mbox{$a(1-e)< 1 <a(1+e)$} originated at (1,0). We note that all lengths have been scaled by the radial distance of the collision point to the Sun, $r_\mathrm{coll}=1$.

An interesting result depicted by figure \ref{dm_dade} is the region where most of the fragments mass is concentrated. As shown here, after an impact, the majority of the mass stays in the vicinity of the collision point in low eccentricity orbits $(a, e)=(1,0)$. To explain this feature, we show in figure \ref{kicks} the $x$ and $y$ components of the velocity of a fragment relative to the velocity of the center of mass of the system immediately after an impact,  \mbox{$\mathbf{v}_\mathrm{rel}=(\mathbf{v}-\mathbf{v}_\mathrm{CoM})/v_\mathrm{CoM}$}. In this equation, $\mathbf{v}$ is the fragment's velocity and $\mathbf{v}_\mathrm{CoM}$ is the velocity of the center of mass. As shown by the top panel of figure \ref{kicks}, the distribution of the relative velocity is nearly spherical and with magnitude generally smaller than 0.3 which confirms the concentration of the mass in the vicinity of the collision point in figure \ref{dm_dade}. As demonstrated by \citet{Jackson2014}, this is consistent with the distribution of fragments that are generated in collisions between low-eccentricity bodies.

Figure \ref{kicks} also shows streaks of high-velocity fragments stretching away from the point of impact. These streaks, also seen in figure \ref{dm_dade}, are produced in high-velocity impacts ($v_\mathrm{0}>2$) on high-eccentric orbits ($e_\mathrm{CoM}>0.33$).
The peculiar fragments distribution produced during these collisions, shown in figure \ref{dm_dade_filtered}, is consistent with the distribution obtained by \citet{Jackson2014} for collisions on high-eccentric orbits and it strongly depends on the eccentricity vector of the two colliding bodies as well as the direction in which the fragments are ejected (streaks in figure \ref{kicks_filtered}).
Even though the fragments distribution produced by these high-velocity and high-eccentric collisions significantly deviates from the overall distribution, the total amount of fragments produced accounts for only $4.1\%$ of the total mass of fragments from the 10 simulations studied.
We therefore decided to keep these collisions in our analysis for completeness.

A detailed inspection of figure \ref{kicks} shows that the center of the velocity distribution lies slightly away from the impact location. While this displacement is negligible along the $x$-axis ($v_{\mathrm{kick},x}=-0.0037$ ), along the $y$-axis, it is rather significant ($v_{\mathrm{kick},y}= -0.036$), indicating a slight loss of angular momentum during the impact. This change in orbital angular momentum, that is partially due to the change in rotation of the main post-collisional bodies and partially due to the loss of energy during the impact, causes the distribution of the fragments to slightly deviate from the collision point and manifests itself in the form of small over-population toward one of the two boundaries. This can be seen in figure \ref{dm_dade} where the left branch ($a<r_\mathrm{coll}$) is slightly more populated than the right branch ($a>r_\mathrm{coll}$).

The slight loss of energy and angular momentum causes the fragments to have slightly smaller semimajor axes. The graph of the cumulative mass distribution of these objects, shown by a solid black curve in Fig. \ref{dm_dade}, show that $64\%$ of the total mass of fragments are in orbits with semimajor axes smaller than the radial position of the collision point $(r_\mathrm{coll})$. In other words, more than half of the fragments have semimajor axes smaller than $0.93\,r_\mathrm{coll}$. The magnitude of this net inward distribution of the post-impact material may not seem substantial for a single collision.
However, when considered in the grand context of the giant impact phase, where there are numerous collision, its cumulative effect in bringing material closer to the central star may be significant.
Furthermore, we investigated the possibility that this effect was produced by a specific class of collisions.
However, we found that, apart from small variations in amplitude, this is not the case.
For example, the high-velocity and high-eccentric collisions shown in figures \ref{dm_dade_filtered} and \ref{kicks_filtered} produce a distribution displacement of $v_{\mathrm{kick},y}= -0.042$.

The bottom panel of figure \ref{kicks} shows the histogram of the post-impact velocity of fragments in terms of their masses. As shown here, for the values of $v_\mathrm{rel}> 0.1$, the decrease in the amplitude of the relative velocity seems to show an exponential trend. For the values of $v_\mathrm{rel}< 0.1$, the distribution deviates from this trend and shows a decrease. This could be attributed to the re-accretion of low-velocity fragments by larger bodies. 
Our simulations indicate that the large majority of the fragments fall in the range of  
0.1< $v_\mathrm{kick}$< 0.6. An exponential fit to this range of the distribution indicates
\begin{equation}\label{kics_fit}
    m(v_\mathrm{kick})\,\mathrm{d}v_\mathrm{kick} \propto \mathrm{e}^{-\tau
      v_\mathrm{kick}}\,\mathrm{d}v_\mathrm{kick}\, ,
\end{equation}
where scaling factor $\tau=11.3\pm0.3$. 

\begin{figure}
  \centering
  \hspace{-0.3cm} 
    \includegraphics[width=0.49\textwidth]{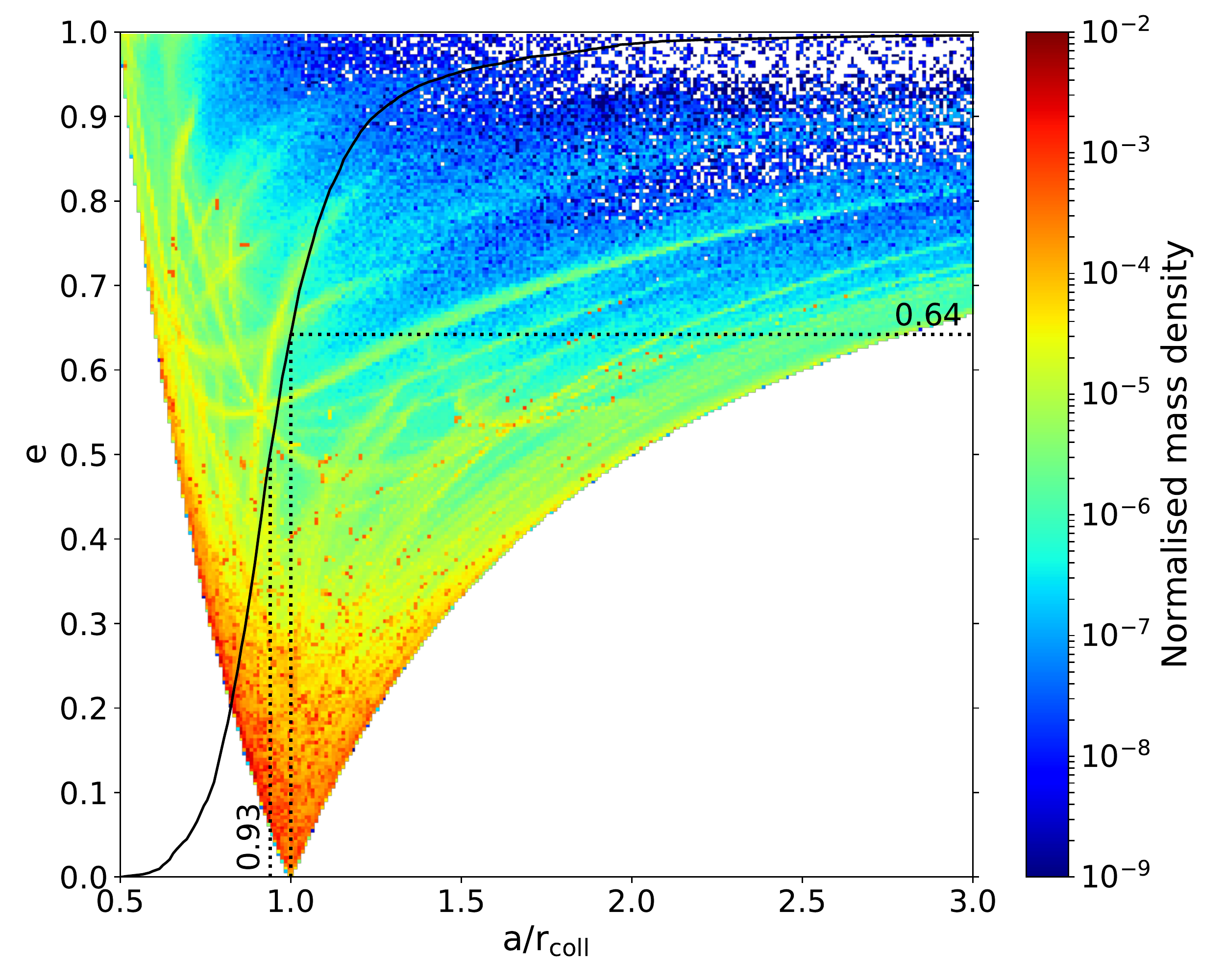}
  \caption{Similar to figure \ref{dm_dade} with Saturn included.}\label{dm_dade_Saturn}
\end{figure}

\vskip 15pt
\noindent
{\it Effect of Saturn}
\vskip 5pt
\noindent
For the sake of comparison, we carried out 10 simulations with the same initial condition as those of our PC catalogue (section 2.1) but included Saturn in a circular orbit. In this case, we obtained 1342 new collisions. Figure \ref{dm_dade_Saturn} shows the distribution of their fragments. A comparison between this figure and figure \ref{dm_dade} shows that the population on the left branch extends to higher values of eccentricities $(\sim 0.8)$, and the distribution with Saturn contains more high-velocity streaks. This could be attributed to Saturn's secular resonance which strongly affects the orbit evolution of planetary embryos \citep{Haghighipour2016}. A comparison with simulations without Saturn showed that the perturbation of Saturn almost doubled the number of high-velocity impacts (i.e., those with $v_\mathrm{0}>4$). As important as the effect of Saturn is, such collisions make a small fraction of the total number of collisions and, statistically, their contribution to the average properties of the ejected material is negligible. For instance, similar to the system of figure \ref{dm_dade}, $64\%$ of the total fragmented mass has semimajor axes smaller than the radial position of collision $r_\mathrm{coll}$.

\subsubsection{Fragments distribution on the $x-y$ plane}\label{sub_sec_x-y_distr}

To construct the distribution of fragments on the $x-y$ plane, we calculated the probability that a fragment with a mass $m_i$ would be in a specific grid-cell $(\Delta x , \Delta y)$. Because fragments move, we weighted this probability by multiplying it by the time ($\Delta t_i$) that a fragment spends inside the cell, modulo the fragment's orbital period $(P_i)$. That is,

\begin{equation}\label{mass_distr}
    \frac{\Delta m}{\Delta x\Delta y}=\sum_i \frac{m_i\, \Delta t_i/P_i}{\Delta x\Delta y}\,.
\end{equation}

\noindent
The distribution was then constructed by summing the contributions of all fragments. Figure \ref{dm_dxdy_distr} shows the results. 

It is important to note that the distribution in Figure \ref{dm_dxdy_distr} did not appear immediately after the collision. As had been reported in other works \citep[e.g.,][]{Jackson2014,Watt}, the post-impact material first formed spiral arms, which in a timescale approximately equal to a hundred Keplerian orbital at $r_\mathrm{coll}$, evolve into the distribution shown in the figure. As expected, fragments distributed around the collision point at (1,0) and gradually spread along the azimuthal angle, almost symmetrical with respect to the line connecting the point of impact to the central star. Further examination of the distribution indicated that in the radial direction, the width of the distribution gradually increased with distance from the impact point with a maximum at the antipode of the collisions point with respect to the central star.

\subsection{Analytical Modeling}

\begin{figure}
  \centering
  \hspace*{-0.5cm} 
    \includegraphics[width=0.54\textwidth]{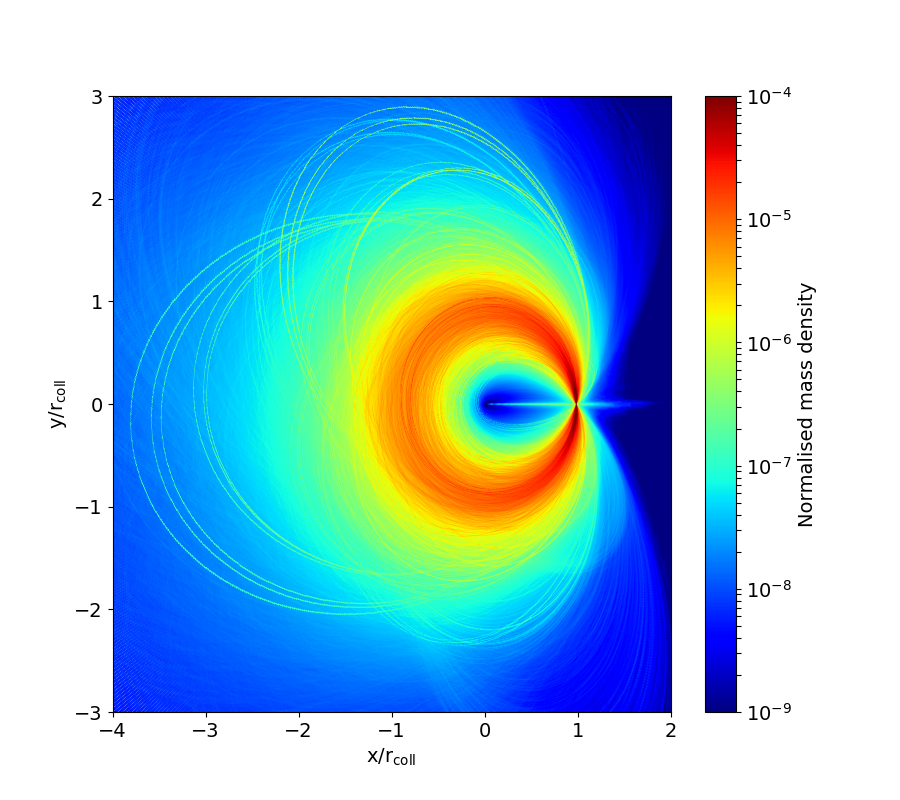}
  \vspace*{-5mm}
  \caption{Spatial distribution of fragments mass in the $x-y$ plane. The collision point is at (1,0).}\label{dm_dxdy_distr}
\end{figure}

As mentioned in the beginning, our goal is to develop a (preferably analytical) model that can be used to demonstrate the outcome of collisions. In this section, we use the distribution shown in figure \ref{dm_dxdy_distr} to develop such a model.

Considering a polar-coordinate system with the central star at the origin and the impact point at $(r=1 , \theta=0)$ (the angular momentum vector pointing towards the reader), we find that the distribution shown in Figure \ref{dm_dxdy_distr} can be modeled by the function
\begin{equation}\label{model}
    \rho\left(r,\theta\right)=
    A\left(\theta\right)\frac{r^{\,\beta\left(\theta\right)}}{\sqrt{1+\left[\frac{r}{r_\mathrm{0}\left(\theta\right)}\right]^{\delta\left(\theta\right)}}}\,.
\end{equation}
where the amplitude $A$ and the exponents $\beta$ and $\delta$ can be obtained using the equation
\begin{equation}\label{f_theta}
    f(\theta)=p_\mathrm{0}-p_\mathrm{1}\cdot\left[\frac{\ln{\left(p_\mathrm{2}+|\theta/\pi-1|\right)}}{p_\mathrm{2}+|\theta/\pi-1|}
    +\frac{\ln{\left(p_\mathrm{2}-|\theta/\pi-1|\right)}}{p_\mathrm{2}-|\theta/\pi-1|}\right]\, .
\end{equation}
The values for the parameters ${p_i}, (i=0-2)$ are given in table \ref{fit_params}, and the length scale $r_\mathrm{0}$ (approximately the maximum in the radial mass distribution for a given value of $\theta$) is given by a sinusoidal function
\begin{equation}\label{r_0}
    r_\mathrm{0}\left(\theta\right)=0.95+0.05\cdot\cos \left(\theta+0.013\right)\,.
\end{equation}
 
\begin{figure}
  \centering
  \hspace*{-0.5cm} 
    \includegraphics[width=0.54\textwidth]{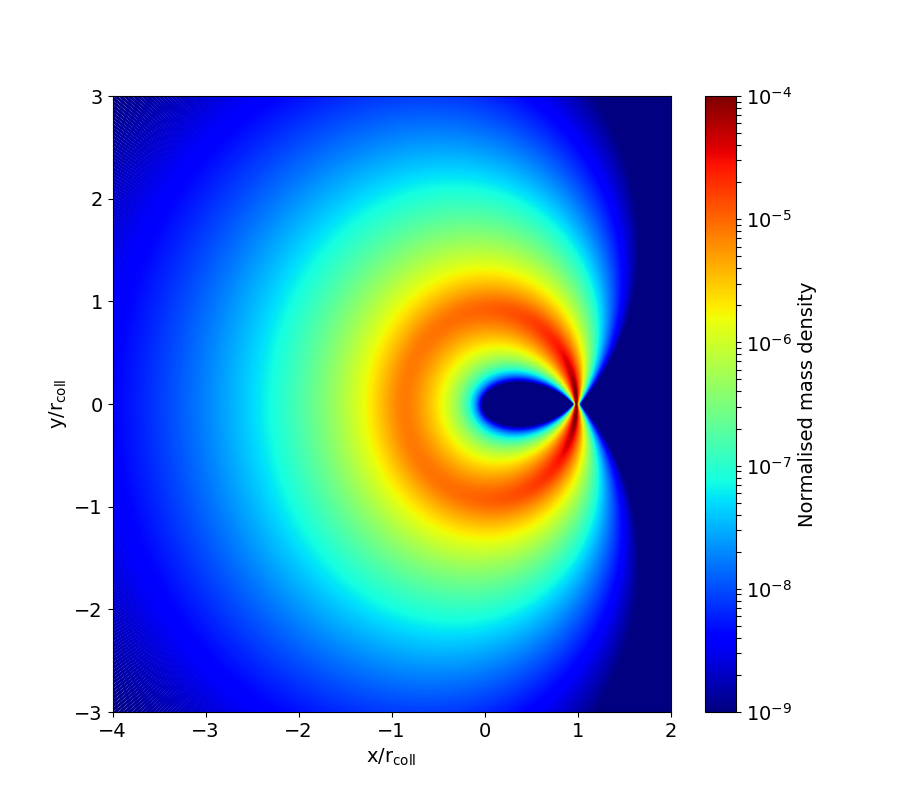}
  \vspace*{-5mm}
  \caption{Model of the mass distribution of the post-impact material of the collision of figure \ref{dm_dxdy_distr} when using equation \ref{model}. 
  }\label{model_dm_dxdy}
\end{figure}

\begin{figure}
  \centering
  \hspace*{-0.3cm} 
    \includegraphics[width=0.49\textwidth]{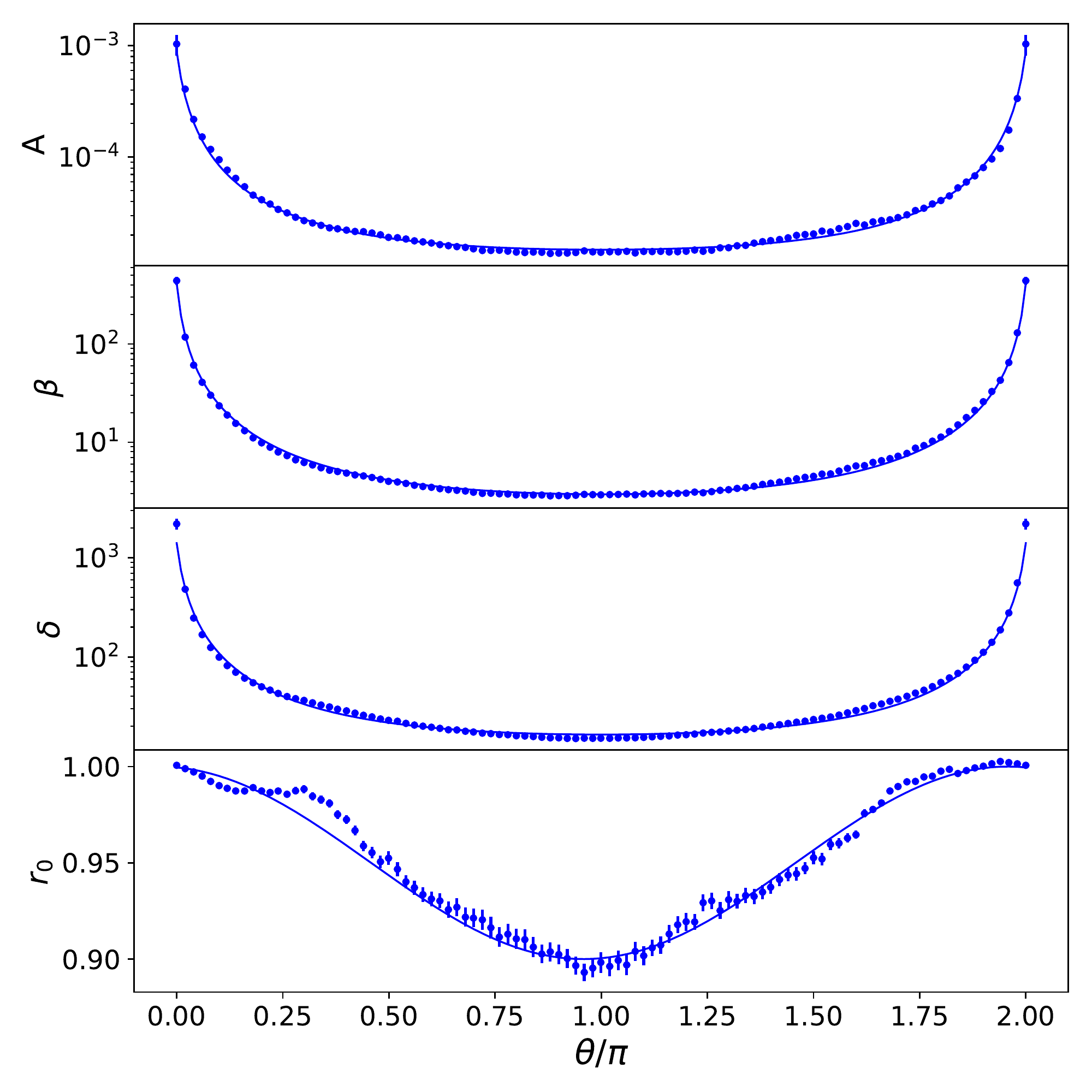}
  \vspace*{-5mm}
  \caption{From top to bottom, graphs of $A$, $\beta$, $\delta$ and $r_\mathrm{0}$ in terms of $\theta$. In all panels, $\theta=0$ represent the position of the impact. The solid curves are the best fits corresponding to equation \ref{f_theta} for $A$, $\beta$, $\delta$ and equation \ref{r_0} for $r_\mathrm{0}$. }\label{model_params}
\end{figure}

Figure \ref{model_dm_dxdy} shows the $x-y$ distribution obtained using equation \ref{model}. A comparison between this figure and figure \ref{dm_dxdy_distr} shows that the model has captured all essential elements of the distribution. For instance, that most of the post-impact material is closer to the central star than the collision point can clearly be seen in this figure. The latter can also be seen in figure \ref{model_params} where the $\theta$-dependence of the four functions $A$, $\beta$, $\delta$ and $r_\mathrm{0}$ are shown. The fact that $r_\mathrm{0}$ oscillates between 1 (at the collision point $\theta=0$) and 0.9 (close to the antipode $\theta=\pi$), indicates that the post-impact material is preferentially scattered into orbits closer to the central star.
Moreover, the minimum of $r_\mathrm{0}$ is directly related to the kick distribution displacement illustrated in section \ref{sub_sec_a-e_distr}. 
In fact, the predominance of kicks in the negative \textit{y}-direction ($v_{\mathrm{kick},y}= -0.036$) causes the argument of periastron for the fragments to center around $\theta=\pi$, while the small displacement of kicks in the negative \textit{x} direction ($v_{\mathrm{kick},x}= -0.0037$) causes the argument of the periastron to slightly move clock-wise inducing a phase of 0.013 in equation \ref{r_0}. 

Lastly, we investigated how the spatial distribution of the fragments is altered when individual subsets of the PC catalogue are considered.
In particular, we divided the collisions into subsets depending on the normalised impact velocity.
We obtained the spatial distribution of fragments for these subsets of collisions using the same analysis performed for the overall spatial distribution (Sec. \ref{sub_sec_x-y_distr}) and we fitted the results using equation \ref{model}.
We obtained that no significant difference can be observed when the spatial distributions from the subsets are compared with the overall distribution in figure \ref{dm_dxdy_distr}.
However, we noticed that the model in equation \ref{model} is extremely sensitive to individual fragments with high mass.
In particular, the length scale $r_\mathrm{0}$ in equation \ref{r_0} tends to match the orbit of the biggest fragments instead of matching the actual maximum of the distribution along the radial direction.
This issue is easily solved when enough collisions are considered, as in the case of the spatial distribution in figure \ref{dm_dxdy_distr}, obtained by considering all the 1356 collisions in our PC catalogue.
Possible future work could increase the size of the PC catalogue to better characterize the spatial distribution of fragments as a function of the collision parameters.

\section{Conclusions}\label{sec:conclusions}

In this study, we present the results of a statistical analysis of the material generated through collisions between protoplanetary bodies during the last stage of the formation of terrestrial planets. Using a catalogue of 1365 impacts and results of the SPH simulations of 880 embryo-embryo collisions, we find that approximately 4.2\% of the total mass of the colliding embryos will be converted into debris. This value can increase up to 24\% \citep{Burger2020} when the masses of the colliding bodies are smaller and a second giant planet is included. 

One important result of our study is the negligible contribution of catastrophic collisions to the total mass of the collisional fragments.
Our simulations show that, in general, the rate of the occurrence of these collisions is small.
For that reason, even though they convert the entire masses of the colliding bodies into debris, they do not contribute much to the total mass of fragments.
This result suggests that when simulating collisions to study the  growth of planetesimals or the late stage of the formation of terrestrial planets, one can ignore catastrophic collisions to a first approximation.

In our simulations we did not consider the distribution of water after each impact. However, volatile materials, such as water, are particularly prone to be lost during these events and to be redistributed across the system \citep{Kegerreis2020}. A sophisticated and robust methodology for including the production of collisional fragments is, therefore, paramount in simulations of planetary formation in general, and in particular in those that aim to track chemicals.

Our analysis indicated that for fragments in the range of $10^{-7}\,M_\oplus$ to $2\times 10^{-2}\,M_\oplus$, the mass distribution can be approximated by an exponential function with an exponent of $-2.21\pm0.17$. Assuming a constant bulk density in this range, this mass distribution translates into an exponential size distribution with an exponent of $-3.63\pm0.52$. This value is consistent with the median value of $3.8$ obtained by \cite{Leinhardt2012}.

Our analysis of the spatial distribution of fragments indicated that the orbits of the large majority of post-impact bodies have eccentricities smaller than 0.4, implying that most of the collisional fragments have low velocities. We also found that many of these fragments $(\sim 64\%)$ are closer to the central star than the point of the impact. This inward scattering of post-impact material is observed for the first time and can be attributed to the loss of angular momentum and energy during an impact. 

Finally, we developed an analytical formula for the distribution of fragments (equation \ref{model}). This equation proves useful in predicting and analyzing results of giant impacts removing the need for carrying out SPH simulations of such events.

\begin{table}
\centering
\begin{tabular}{cccc}
    $f(\theta)$ & $p_\mathrm{0}$ & $p_\mathrm{1}$ & $p_\mathrm{2}$  \\ \hline
    $A(\theta)$ & $1.49\cdot 10^{-5}$ & $3.89\cdot 10^{-6}$ & $1.018$\\
    $\beta(\theta)$ & $2.98$ & $1.10$ & $1.012$\\
    $\delta(\theta)$ & $16.6$ & $4.99$ & $1.015$\\
\end{tabular}
\caption{Bet fit parameters for the functions $A$, $\beta$ and $\delta$ with \mbox{equation \ref{f_theta}} as fit function.}
\label{fit_params}
\end{table}

\section*{Acknowledgements}

We are thankful to the referee, John Chambers, for his critically reading of our paper and his useful comments that improved our manuscript.
This work was supported by the NYU Abu Dhabi Global Ph.D. Student Fellowship Program.
TIM acknowledges the support from the Austrian Science Fund (FWF): P33351-N.
CMS appreciates support by the DFG German Science Foundation projects 398488521, 446102036, 285676328. NH acknowledges support from NASA grant 80NSSC21K1050.

The following software facilitate the advancement of this work: Matplotlib \citep{Hunter2007}, NumPy \citep{vanderwalt2011}.

\section*{Data Availability}

The data underlying this article will be shared on reasonable request to the corresponding author.



\bibliographystyle{mnras}

\begin{thebibliography}{}
\makeatletter
\relax
\def\mn@urlcharsother{\let\do\@makeother \do\$\do\&\do\#\do\^\do\_\do\%\do\~}
\def\mn@doi{\begingroup\mn@urlcharsother \@ifnextchar [ {\mn@doi@}
  {\mn@doi@[]}}
\def\mn@doi@[#1]#2{\def\@tempa{#1}\ifx\@tempa\@empty \href
  {http://dx.doi.org/#2} {doi:#2}\else \href {http://dx.doi.org/#2} {#1}\fi
  \endgroup}
\def\mn@eprint#1#2{\mn@eprint@#1:#2::\@nil}
\def\mn@eprint@arXiv#1{\href {http://arxiv.org/abs/#1} {{\tt arXiv:#1}}}
\def\mn@eprint@dblp#1{\href {http://dblp.uni-trier.de/rec/bibtex/#1.xml}
  {dblp:#1}}
\def\mn@eprint@#1:#2:#3:#4\@nil{\def\@tempa {#1}\def\@tempb {#2}\def\@tempc
  {#3}\ifx \@tempc \@empty \let \@tempc \@tempb \let \@tempb \@tempa \fi \ifx
  \@tempb \@empty \def\@tempb {arXiv}\fi \@ifundefined
  {mn@eprint@\@tempb}{\@tempb:\@tempc}{\expandafter \expandafter \csname
  mn@eprint@\@tempb\endcsname \expandafter{\@tempc}}}

\bibitem[\protect\citeauthoryear{{Barnes}, {Quinn}, {Lissauer}  \&
  {Richardson}}{{Barnes} et~al.}{2009}]{Barnes2009}
{Barnes} R.,  {Quinn} T.~R.,  {Lissauer} J.~J.,   {Richardson} D.~C.,  2009,
  \mn@doi [\icarus] {10.1016/j.icarus.2009.03.042}, \href
  {https://ui.adsabs.harvard.edu/abs/2009Icar..203..626B} {203, 626}

\bibitem[\protect\citeauthoryear{{Benz} \& {Asphaug}}{{Benz} \&
  {Asphaug}}{1994}]{Benz1994}
{Benz} W.,  {Asphaug} E.,  1994, \mn@doi [Icarus] {10.1006/icar.1994.1009},
  \href {https://ui.adsabs.harvard.edu/abs/1994Icar..107...98B} {107, 98}

\bibitem[\protect\citeauthoryear{{Benz}, {Slattery}  \& {Cameron}}{{Benz}
  et~al.}{1986}]{Benz1986}
{Benz} W.,  {Slattery} W.~L.,   {Cameron} A.~G.~W.,  1986, \mn@doi [\icarus]
  {10.1016/0019-1035(86)90088-6}, \href
  {https://ui.adsabs.harvard.edu/abs/1986Icar...66..515B} {66, 515}

\bibitem[\protect\citeauthoryear{{Burger}, {Maindl}  \& {Sch{\"a}fer}}{{Burger}
  et~al.}{2018}]{Burger2018}
{Burger} C.,  {Maindl} T.~I.,   {Sch{\"a}fer} C.~M.,  2018, \mn@doi [Celestial
  Mechanics and Dynamical Astronomy] {10.1007/s10569-017-9795-3}, \href
  {https://ui.adsabs.harvard.edu/abs/2018CeMDA.130....2B} {130, 2}

\bibitem[\protect\citeauthoryear{{Burger}, {Bazs{\'o}}  \&
  {Sch{\"a}fer}}{{Burger} et~al.}{2020}]{Burger2020}
{Burger} C.,  {Bazs{\'o}} {\'A}.,   {Sch{\"a}fer} C.~M.,  2020, \mn@doi [\aap]
  {10.1051/0004-6361/201936366}, \href
  {https://ui.adsabs.harvard.edu/abs/2020A&A...634A..76B} {634, A76}

\bibitem[\protect\citeauthoryear{{Carter}, {Leinhardt}, {Elliott}, {Walter}  \&
  {Stewart}}{{Carter} et~al.}{2015}]{Carter2015}
{Carter} P.~J.,  {Leinhardt} Z.~M.,  {Elliott} T.,  {Walter} M.~J.,   {Stewart}
  S.~T.,  2015, \mn@doi [\apj] {10.1088/0004-637X/813/1/72}, \href
  {https://ui.adsabs.harvard.edu/abs/2015ApJ...813...72C} {813, 72}

\bibitem[\protect\citeauthoryear{{Chambers}}{{Chambers}}{1999}]{Chambers1999}
{Chambers} J.~E.,  1999, \mn@doi [\mnras] {10.1046/j.1365-8711.1999.02379.x},
  \href {https://ui.adsabs.harvard.edu/abs/1999MNRAS.304..793C} {304, 793}

\bibitem[\protect\citeauthoryear{{Chambers}}{{Chambers}}{2013}]{Chambers2013}
{Chambers} J.~E.,  2013, \mn@doi [\icarus] {10.1016/j.icarus.2013.02.015},
  \href {https://ui.adsabs.harvard.edu/abs/2013Icar..224...43C} {224, 43}

\bibitem[\protect\citeauthoryear{{Clement}, {Kaib}, {Raymond}, {Chambers}  \&
  {Walsh}}{{Clement} et~al.}{2019}]{Clement2019}
{Clement} M.~S.,  {Kaib} N.~A.,  {Raymond} S.~N.,  {Chambers} J.~E.,   {Walsh}
  K.~J.,  2019, \mn@doi [\icarus] {10.1016/j.icarus.2018.12.033}, \href
  {https://ui.adsabs.harvard.edu/abs/2019Icar..321..778C} {321, 778}

\bibitem[\protect\citeauthoryear{{Dormand} \& {Woolfson}}{{Dormand} \&
  {Woolfson}}{1977}]{1977Dormand}
{Dormand} J.~R.,  {Woolfson} M.~M.,  1977, \mn@doi [\mnras]
  {10.1093/mnras/180.2.243}, \href
  {https://ui.adsabs.harvard.edu/abs/1977MNRAS.180..243D} {180, 243}

\bibitem[\protect\citeauthoryear{{Dugaro}, {de El{\'\i}a}  \&
  {Darriba}}{{Dugaro} et~al.}{2020}]{Doguaro2020}
{Dugaro} A.,  {de El{\'\i}a} G.~C.,   {Darriba} L.~A.,  2020, \mn@doi [\aap]
  {10.1051/0004-6361/202037619}, \href
  {https://ui.adsabs.harvard.edu/abs/2020A&A...641A.139D} {641, A139}

\bibitem[\protect\citeauthoryear{{Haghighipour} \& {Winter}}{{Haghighipour} \&
  {Winter}}{2016}]{Haghighipour2016}
{Haghighipour} N.,  {Winter} O.~C.,  2016, \mn@doi [Celestial Mechanics and
  Dynamical Astronomy] {10.1007/s10569-015-9663-y}, \href
  {https://ui.adsabs.harvard.edu/abs/2016CeMDA.124..235H} {124, 235}

\bibitem[\protect\citeauthoryear{{Hunter}}{{Hunter}}{2007}]{Hunter2007}
{Hunter} J.~D.,  2007, \mn@doi [Computing in Science Engineering]
  {10.1109/MCSE.2007.55}, 9, 90

\bibitem[\protect\citeauthoryear{{Ida} \& {Makino}}{{Ida} \&
  {Makino}}{1993}]{Ida1993}
{Ida} S.,  {Makino} J.,  1993, \mn@doi [\icarus] {10.1006/icar.1993.1167},
  \href {https://ui.adsabs.harvard.edu/abs/1993Icar..106..210I} {106, 210}

\bibitem[\protect\citeauthoryear{{Jackson}, {Wyatt}, {Bonsor}  \&
  {Veras}}{{Jackson} et~al.}{2014}]{Jackson2014}
{Jackson} A.~P.,  {Wyatt} M.~C.,  {Bonsor} A.,   {Veras} D.,  2014, \mn@doi
  [\mnras] {10.1093/mnras/stu476}, \href
  {https://ui.adsabs.harvard.edu/abs/2014MNRAS.440.3757J} {440, 3757}

\bibitem[\protect\citeauthoryear{{Kegerreis}, {Eke}, {Catling}, {Massey},
  {Teodoro}  \& {Zahnle}}{{Kegerreis} et~al.}{2020}]{Kegerreis2020}
{Kegerreis} J.~A.,  {Eke} V.~R.,  {Catling} D.~C.,  {Massey} R.~J.,  {Teodoro}
  L.~F.~A.,   {Zahnle} K.~J.,  2020, \mn@doi [\apjl]
  {10.3847/2041-8213/abb5fb}, \href
  {https://ui.adsabs.harvard.edu/abs/2020ApJ...901L..31K} {901, L31}

\bibitem[\protect\citeauthoryear{{Kokubo} \& {Ida}}{{Kokubo} \&
  {Ida}}{2000}]{Kokubo2000}
{Kokubo} E.,  {Ida} S.,  2000, \mn@doi [\icarus] {10.1006/icar.1999.6237},
  \href {https://ui.adsabs.harvard.edu/abs/2000Icar..143...15K} {143, 15}

\bibitem[\protect\citeauthoryear{{Leinhardt} \& {Stewart}}{{Leinhardt} \&
  {Stewart}}{2012}]{Leinhardt2012}
{Leinhardt} Z.~M.,  {Stewart} S.~T.,  2012, \mn@doi [\apj]
  {10.1088/0004-637X/745/1/79}, \href
  {https://ui.adsabs.harvard.edu/abs/2012ApJ...745...79L} {745, 79}

\bibitem[\protect\citeauthoryear{Monaghan \& Pongracic}{Monaghan \&
  Pongracic}{1985}]{Monaghan1985}
Monaghan J.,  Pongracic H.,  1985, \mn@doi [Applied Numerical Mathematics]
  {https://doi.org/10.1016/0168-9274(85)90015-7}, 1, 187

\bibitem[\protect\citeauthoryear{{Morishima}}{{Morishima}}{2015}]{Morishima2015}
{Morishima} R.,  2015, \mn@doi [\icarus] {10.1016/j.icarus.2015.07.030}, \href
  {https://ui.adsabs.harvard.edu/abs/2015Icar..260..368M} {260, 368}

\bibitem[\protect\citeauthoryear{{Mustill}, {Davies}  \& {Johansen}}{{Mustill}
  et~al.}{2018}]{Mustill2018}
{Mustill} A.~J.,  {Davies} M.~B.,   {Johansen} A.,  2018, \mn@doi [\mnras]
  {10.1093/mnras/sty1273}, \href
  {https://ui.adsabs.harvard.edu/abs/2018MNRAS.478.2896M} {478, 2896}

\bibitem[\protect\citeauthoryear{{O'Brien}, {Morbidelli}  \&
  {Levison}}{{O'Brien} et~al.}{2006}]{OBrien2006}
{O'Brien} D.~P.,  {Morbidelli} A.,   {Levison} H.~F.,  2006, \mn@doi [\icarus]
  {10.1016/j.icarus.2006.04.005}, \href
  {https://ui.adsabs.harvard.edu/abs/2006Icar..184...39O} {184, 39}

\bibitem[\protect\citeauthoryear{{Poon}, {Nelson}, {Jacobson}  \&
  {Morbidelli}}{{Poon} et~al.}{2020}]{Poon2020}
{Poon} S. T.~S.,  {Nelson} R.~P.,  {Jacobson} S.~A.,   {Morbidelli} A.,  2020,
  \mn@doi [\mnras] {10.1093/mnras/stz3296}, \href
  {https://ui.adsabs.harvard.edu/abs/2020MNRAS.491.5595P} {491, 5595}

\bibitem[\protect\citeauthoryear{{Raymond}, {Quinn}  \& {Lunine}}{{Raymond}
  et~al.}{2004}]{Raymond2004}
{Raymond} S.~N.,  {Quinn} T.,   {Lunine} J.~I.,  2004, \mn@doi [\icarus]
  {10.1016/j.icarus.2003.11.019}, \href
  {https://ui.adsabs.harvard.edu/abs/2004Icar..168....1R} {168, 1}

\bibitem[\protect\citeauthoryear{{Raymond}, {Quinn}  \& {Lunine}}{{Raymond}
  et~al.}{2006}]{Raymond2006}
{Raymond} S.~N.,  {Quinn} T.,   {Lunine} J.~I.,  2006, \mn@doi [\icarus]
  {10.1016/j.icarus.2006.03.011}, \href
  {https://ui.adsabs.harvard.edu/abs/2006Icar..183..265R} {183, 265}

\bibitem[\protect\citeauthoryear{{Sch{\"a}fer}, {Riecker}, {Maindl}, {Speith},
  {Scherrer}  \& {Kley}}{{Sch{\"a}fer} et~al.}{2016}]{Schafer2016}
{Sch{\"a}fer} C.,  {Riecker} S.,  {Maindl} T.~I.,  {Speith} R.,  {Scherrer} S.,
    {Kley} W.,  2016, \mn@doi [Astronomy \& Astrophysics]
  {10.1051/0004-6361/201528060}, \href
  {https://ui.adsabs.harvard.edu/abs/2016A&A...590A..19S} {590, A19}

\bibitem[\protect\citeauthoryear{{Sch{\"a}fer} et~al.,}{{Sch{\"a}fer}
  et~al.}{2020}]{Schafer2020}
{Sch{\"a}fer} C.~M.,  et~al., 2020, \mn@doi [Astronomy and Computing]
  {10.1016/j.ascom.2020.100410}, \href
  {https://ui.adsabs.harvard.edu/abs/2020A&C....3300410S} {33, 100410}

\bibitem[\protect\citeauthoryear{{Scora}, {Valencia}, {Morbidelli}  \&
  {Jacobson}}{{Scora} et~al.}{2020}]{Scora2020}
{Scora} J.,  {Valencia} D.,  {Morbidelli} A.,   {Jacobson} S.,  2020, \mn@doi
  [\mnras] {10.1093/mnras/staa568}, \href
  {https://ui.adsabs.harvard.edu/abs/2020MNRAS.493.4910S} {493, 4910}

\bibitem[\protect\citeauthoryear{Van Der~Walt, Colbert  \& Varoquaux}{Van
  Der~Walt et~al.}{2011}]{vanderwalt2011}
Van Der~Walt S.,  Colbert S.~C.,   Varoquaux G.,  2011, \mn@doi [{Computing in
  Science and Engineering}] {10.1109/MCSE.2011.37}, 13, 22

\bibitem[\protect\citeauthoryear{Watt, Leinhardt  \& Su}{Watt
  et~al.}{2021}]{Watt}
Watt L.,  Leinhardt Z.,   Su K.,  2021, \mn@doi [Monthly Notices of the Royal
  Astronomical Society] {10.1093/mnras/stab106}

\bibitem[\protect\citeauthoryear{{Wetherill}}{{Wetherill}}{1988}]{Wetherill1988}
{Wetherill} G.~W.,  1988, {Accumulation of Mercury from planetesimals}.
pp 670--691

\makeatother
\end{thebibliography}







\bsp	
\label{lastpage}
\end{document}